\documentclass[useAMS,usenatbib]{mnras}
\usepackage{tabularx}
\usepackage{url}
\usepackage{graphicx}
\usepackage[usenames]{color}
\usepackage[T1]{fontenc}
\usepackage{ae,aecompl}
\usepackage{booktabs}
\def\unit #1{\,{\rm #1}}

\newcommand\kms{\rm \,\unit{km\,s^{-1}}}

\newcommand\cmsqi{\rm \,\unit{cm^{-2}}}

\newcommand\s{\rm \,\unit{s}}

\newcommand\kev{\rm \,\unit{keV}}

\newcommand\ergs{\rm \,\unit{erg\,s^{-1}}}

\newcommand\funit{\rm \,erg\,cm^{-2}\,s^{-1}}

\newcommand\lunit{\rm \,erg \,s^{-1}}

\newcommand\xiunit{\rm \,erg\,cm\,s^{-1}}

\newcommand\msol{M_{\odot}}

\newcommand\ks{\, \rm ks}

\newcommand\dc{\, \Delta\chi^2}

\newcommand\cd{\,\rm \chi^2/dof}

\newcommand\kpc{\unit{kpc}}

\newcommand\mpc{\unit{Mpc}}

\newcommand\ev{\unit{\, eV}}

\def\ltsim{\mathrel{\hbox{\rlap{\hbox{\lower3pt\hbox{$\sim$}}}\hbox{\raise2pt\hbox{$<$}}}}}
\def\gtrsim{\mathrel{\hbox{\rlap{\hbox{\lower3pt\hbox{$\sim$}}}\hbox{\raise2pt\hbox{$>$}}}}}

\newcommand\fermi{{\it FERMI}}
\newcommand\suzaku{{\it Suzaku}}
\newcommand\nustar{{\it NuSTAR}}

\newcommand\swift{{\it Swift}}

\def\pasp{PASP} 
\def\apj{ApJ} \def\mnras{MNRAS}
\def\aap{A\&A} \def\apjl{ApJ} \def\aj{aj} \def\physrep{PhR}
 \def\apjs{ApJS}  \def\pasj{PASJ}
\def\nat{Nat} \def\ssr{SSRv}

\title[Broadband spectral study of 1H~0323+342]{Broadband spectral study of the jet-disc emission in the radio-loud narrow-line Seyfert 1 galaxy 1H~0323+342}

\author [Ghosh, Dewangan, Mallick \& Raychaudhuri] {Ritesh Ghosh$^{1}$\thanks{Email: riteshghosh.rs@visva-bharati.ac.in}, 
Gulab C.\ Dewangan$^{2}$\thanks{Email: gulabd@iucaa.in}, Labani Mallick$^{2}$\thanks{Email: labani@iucaa.in}, Biplab Raychaudhuri$^{1}$\thanks{Email: biplabphy@visva-bharati.ac.in} \\
$^{1}$ Visva-Bharati University, Santiniketan, India \\ 
$^{2}$ Inter-University Centre for Astronomy and Astrophysics (IUCAA), Pune, India}

\begin{document}
\maketitle


\begin{abstract}

We present a broadband spectral study of the radio-loud narrow-line Seyfert 1 galaxy 1H~0323+342 based on multi-epoch observations performed with \nustar{} on 2014 March 15, and two simultaneous observations performed with \suzaku{} and \swift{} on 2009 July 26 and 2013 March 1. We found the presence of a strong soft X-ray excess emission, a broad but weak Fe line and hard X-ray excess emission. We used the blurred reflection ({\tt relxill}) and the intrinsic disc Comptonization ({\tt optxagnf}), two physically motivated models, to describe the broadband spectra and to disentangle the disk/corona and jet emission. The {\tt relxill} model is mainly constrained by the strong soft X-ray excess although the model failed to predict this excess when fitted above $3\kev$ and extrapolated to lower energies. The joint spectral analysis of the three datasets above $3\kev$ with this model resulted in a high black hole spin ($a>0.9$) and moderate reflection fraction $R\sim 0.5$. The {\tt optxagnf} model fitted to the two simultaneous datasets resulted in an excess emission in the UV band. The simultaneous UV-to-hard X-ray spectra of 1H~0323+342 are best described by a model consisting of a primary X-ray power-law continuum with $\Gamma \sim 1.8$, a blurred reflection component with $R\sim 0.5$, Comptonised disk emission as the soft X-ray excess, optical/UV emission from a standard accretion disk around a black hole of mass $\sim 10^7{\rm M_{\odot}}$ and a steep power law ($\Gamma \sim 3-3.5$) component, most likely the jet emission in the UV band. The fractional RMS variability spectra 
suggest that both the soft excess and the powerlaw component are variable in nature.

\end{abstract}

\begin{keywords}

accretion, accretion disks - galaxies: active - galaxies: individual (1H~0323+342) - galaxies: Seyfert - X-rays: galaxies 
  
\end{keywords}



\section{Introduction}

The radio-loud active galactic nuclei (AGN) with relativistic jets emitting powerful radio emission constitute a small fraction of all AGN \citep{1989AJ.....98.1195K,1995PASP..107..803U,2002AJ....124.2364I}.
Generally $15\%-20\%$ of  AGN show radio-loudness~\citep{1995PASP..107..803U}, but  only 7\% of narrow-line Seyfert-1 galaxies (NLS1s) are found to be formally radio-loud and  only 2.5\% NLS1s show  strong radio loudness $R_{5GHz} > 100$ \citep{2006AJ....132..531K}.  
The NLS1s are usually found to possess lower mass black holes and higher accretion rates compared to the broad-line Seyfert 1 galaxies.
\citet{2008ApJ...685..801Y} revealed 23 genuine radio-loud NLS1s with  $R_{1.4GHz} > 100$. Interestingly, their observational properties were similar to the  blazars with relativistic jets. Although this finding has been confirmed and supported by the \fermi{}/{\rm LAT} detections of gamma-ray emission in the GeV band~\citep{2009ApJ...707.1310A,2009ApJ...707L.142A}, there are some notable differences. Blazars are generally believed to produce gamma-ray emission from jets powered by the spin-down of the ergosphere of a rotating supermassive black hole. The accretion rates in blazars appear to be far below the Eddington limit \citep{2010MNRAS.402..497G}. This scenario contradicts that of the NLS1 galaxies which are believed to host smaller central black hole masses and accrete at rates close to the Eddington limit~\citep{1992ApJS...80..109B,2004ApJ...606L..41G,2012AJ....143...83X}.

According to the Blandford-Znajek mechanism the production of powerful radio jets requires a highly spinning black hole system which in turn indicates the last stable orbit approaching much closer to the black hole than that in the Schwarzschild geometry.
Thus, emission from inner accretion disk/corona such as that possible in NLS1 galaxies may play a crucial role in understanding the conditions giving rise to the jets.
The RLNLS1s with strong jet and disk/corona emission and shorter variability timescales may help in a phenomenological understanding of the disc-jet connection in AGN which, unlike X-ray Binaries, is not well established to date. 

Recent studies of RLAGN e.g. \citet{2009ApJ...704.1689C,2011MNRAS.418L..89T}, present a scenario where both the Galactic microquasars and luminous radio-loud AGN display complex ``jet cycles'' linking the jet and the disc\citep{2013ApJ...772...83L}. Since the relativistic X-ray reflection, especially the broad iron line offers a powerful tool to investigate the inner regions of AGN accretion disks~\citep{1995Natur.375..659T,1995MNRAS.277L..11F,2003PhR...377..389R}, measuring the inner extent of the accretion disk during several phases of the jet cycle is likely to provide useful information to boost this scenario. Thus, the class of AGN with strong jet and disc/corona emission are the most suitable for the study of disc-jet connection.

The RLNLS1 or NLS1-Blazar 1H~0323+342 is an AGN with relativistic jets as revealed by the  gamma-ray emission discovered with \fermi{}, and strong accretion disk emission as revealed by broad optical/UV emission lines requiring the ionising continuum from the disk.  It is a nearby (z=0.061) NLS1 with strong radio ($R_{1.4GHz} = 318$; \citealt{2011nlsg.confE....F}) and hard X-ray emission. It is known to harbour a black hole with a mass of $\sim 10^{7}\msol$ and has a hybrid nature showing properties of both the NLS1s and the blazars~\citep{2007ApJ...658L..13Z}. The VLA 1.4GHz image revealed a two-sided structure of $\sim 15\kpc$ around the core~\citep{2008A&A...490..583A}. The $0.3-10\kev$ \swift{} XRT data were fitted with a power law of $\Gamma \sim 2$~\citep{2007ApJ...658L..13Z} and the spectrum is found to flatten above $20\kev$ up to $100\kev$ with $\Gamma = 1.55$. The X-ray spectrum was inferred to be reflection dominated by \citet{2014ApJ...789..143P} and \citet{2013MNRAS.428.2901W} as they found the presence of a Fe emission line and a high value of black hole spin. These results, however, are contradicted by the findings of \citet{2015AJ....150...23Y} and thus remain ambiguous. Previously, \fermi{}/{\rm LAT} detection of 1H~0323+342 found the presence of jet emission in the broadband spectral energy distribution in addition to an accretion disk component~\citep{2009ApJ...707L.142A}. This may justify the two bumps, one at the radio-infrared band and the other at the GeV gamma-ray band. The hard X-ray flux variation above the $20\kev$ was found by \cite{2009AdSpR..43..889F}, and the low and the high flux were interpreted as emissions from the disk/corona and the jet flaring, respectively. 
Since the spectrum in the hard X-ray band is similar to a blazar, its X-ray spectrum seems to possess different components originated from the disk and the jet. Since beamed radiation from the jet varies to a great extent, the
temporal behaviour is also important to differentiate between the emission components, e.g., the jet and the disc/corona emission. The source 1H~0323+342 is relatively close and  an excellent target to investigate the broadband emission from the disc and the jet which also allows us to study the possible jet cycle and/or coronal outflow feature found in some radio-loud AGNs.

In this paper, we have studied two sets of simultaneous, broadband, high-quality data from \suzaku{} and \swift{} and a single \nustar{} observation and used physically motivated models to disentangle the disk/corona and jet emission from this gamma-ray loud RLNLS1 AGN.
The paper is organized as follows. In Section 2, we describe the observations and data reduction techniques. We describe the spectral fitting of individual datasets as well as joint spectral analysis in Section 3. In Section 4, we report the fractional RMS spectral analysis. Finally, in Section 5, we discuss our results and summarize our findings. Throughout this paper, we assumed a cosmology with $H_{0} = 71\kms \mpc^{-1}, \Omega_{\Lambda} = 
0.73$ and $\Omega_{M} = 0.27$.

\section{observations and Data Reduction}

\subsection{\suzaku{} observations}
\suzaku{} observed the source 1H~0323+342 twice on 2009 July 26~\citep{2015AJ....150...23Y,2013MNRAS.428.2901W} and 2013 March 1 using the onboard X-ray
Imaging Spectrometer (XIS; \citealt{2007PASJ...59S..23K}) and the Hard X-ray Detector (HXD; \citealt{2007PASJ...59S..35T}) (Table~\ref{observation Table}). The three XIS (the XIS0 and the XIS3 are front-illuminated while the XIS1 is back-illuminated) CCD cameras and the HXD/PIN detector cover the energy range $0.2-12 \kev$ and $10-70\kev$ respectively. The $3 \times 3$ and $5 \times 5$ data modes were used to obtain the XIS data for both of the observations. We 
used the version 6.19 of {\tt HEASOFT} software and the corresponding calibration files to process the \suzaku{} data. The {\sc aepipeline} tool was used to reprocess and clean the unfiltered event files. We extracted the source spectrum for each observation from the filtered event lists and a $240{\rm~arcsec}$ circular region around the source position was used for all the three XIS cameras. The background spectral data were extracted using four circular regions of $120{\rm~arcsec}$ radii, excluding the
source region. The Suzaku tools the {\tt xissimarfgen} and the {\tt xisrmfgen} were used to produce the ``ARF'' and the ``RMF'' files for each XIS spectral data. The {\sc aepipeline} tool also generated the PIN cleaned events and the pseudo-event lists. The {\sc hxdpinxbpi} tool was used to extract the hard X-ray spectral data using these files. The background estimation for the non-imaging HXD/PIN data required both of the non-X-ray instrumental background (NXB) and the cosmic X-ray background (CXB). These background estimations were provided by the \suzaku{} team. We downloaded the corresponding tuned background files available at the {\sc HEASARC}
website~\footnote{http://heasarc.gsfc.nasa.gov/docs/suzaku/analysis/pinbgd.html}. The XIS and the PIN datasets were grouped to a minimum of 200 and 20 counts in each energy bin respectively.

\subsection{\swift{} observations}
1H~0323+342 has been observed simultaneously with the \swift{} mission~\citep{2004ApJ...611.1005G} on both 2009 and 2013~\citep{2015AJ....150...23Y}. Simultaneous \swift{} data enable us to study the optical-to-hard X-ray spectrum of 1H~0323+342. The Ultraviolet-Optical Telescope (UVOT;~\citealt{2005SSRv..120...95R}) observed the source 1H~0323+342 in all the six filters i.e. in the optical ($V, B, U$) bands and the near
UV ($W1, M2, W2$) bands. Co-added image files were used to calculate the count rates and we used circular regions of radius $5^{''}$ and $20^{''}$ for the source and background respectively. Both source and background pha files were created using the UVOT2PHA tool. The corresponding response files were provided by the \swift{} team~\citep{2008MNRAS.383..627P,2010MNRAS.406.1687B}. This source is repeatedly observed by the Burst Alert Telescope (BAT,~\citealt{2005SSRv..120..143B}) and included in the Swift-BAT 70-month hard X-ray catalogue~\citep{2013ApJS..207...19B}. We downloaded the publicly available 70-month averaged spectrum of 1H~0323+342 and compared it with our \nustar{} and \suzaku{} hard X-ray data.

\subsection{\nustar{} observation}
1H~0323+342 was observed with the Nuclear Spectroscopic Telescope Array (\nustar{};~\citealt{2013ApJ...770..103H}) on 2014 March 15, with an exposure time of 101 ks. We followed standard procedures and used $NuSTARDAS v1.6.0$ (included in {\tt HEASOFT, version 6.19}) and \nustar{} CALDB 20170222 to extract the spectral products. Source data from both focal plane modules (FPMA and FPMB) were extracted using a circular region of radius $60^{''}$ centred on the source position of 1H~0323+342. An $80^{''}$ radius source-free region of the same chip of the detector plane was used to extract the background. The NuSTAR datasets were grouped by a minimum of 100 counts per energy bin.

\begin{table}
\centering
\footnotesize
  \caption{Details of observations used in this work. \label{observation Table}}

{\renewcommand{\arraystretch}{1.5}
\setlength{\tabcolsep}{2pt}
\begin{tabular}{ccccc} \hline
   Telescope   & obsID 	  & Date of      &  Exposure   &  Count rate   \\
               &          & observation  &   time      &  (${\rm counts~s^{-1}}$)\\ \hline
  \suzaku{} &704034010  &2009/07/26         & 84ks (XIS)  & $0.66^{+0.03}_{-0.03}$\\
            &           & (Obs 1)           & 74ks (HXD)  & $0.02^{+0.01}_{-0.01}$ \\
  \suzaku{} &707015010  &2013/03/01         & 101ks (XIS) & $0.65^{+0.03}_{-0.03}$\\
            &           & (Obs 2)           &  91ks (HXD) & $0.06^{+0.01}_{-0.01}$ \\
 \nustar{} &60061360002 &2014/03/15         & 91ks (FPMA) & $0.24^{+0.01}_{-0.01}$\\
           &            & (Obs 3)           & 90ks (FPMB) & $0.25^{+0.01}_{-0.01}$\\ 
  \swift{}     &00036533014  &2009/07/27    &237s (V) &$8.61^{+0.30}_{-0.30}$ \\
               &             &              &237s (B) &$1.49^{+0.41}_{-0.41}$ \\
               &             &              &237s (U) &$1.68^{+0.47}_{-0.47}$ \\
               &             &              &475s (UVW1) &$5.80^{+0.16}_{-0.16}$ \\
               &             &              &951s (UVW2) &$4.50^{+0.12}_{-0.12}$ \\
               &             &              &569s (UVM2) &$2.42^{+0.09}_{-0.09}$ \\
               &00036533036  &2013/03/02 &99s  (V) & $7.65^{+0.39}_{-0.39}$\\ 
               &             &              &189s (B) &$1.60^{+0.49}_{-0.49}$ \\
               &             &              &189s (U) &$1.60^{+0.49}_{-0.49}$ \\
               &             &              &378s (UVW1) &$4.92^{+0.16}_{-0.16}$ \\
               &             &              &670s (UVW2) &$3.80^{+0.11}_{-0.11}$ \\
               &             &              &325s (UVM2) &$2.06^{+0.10}_{-0.10}$ \\ \hline\hline

\end{tabular}}
Note: The exposure times quoted are after filtering the data sets.
\end{table}

\section{Spectral analysis and Results}

 We used {\sc XSPEC}~\citep{1996ASPC..101...17A} version 12.9.0n, and employed a $\chi^{2}$ minimization technique for all our spectral analysis. We quote the errors on the best-fit 
parameters  at the $90\%$ confidence level.
To enhance the signal to noise ratio, we added the spectral data from the front-illuminated CCDs of \suzaku{} (the XIS0 and the XIS3)  using the {\sc addascaspec} tool for both \suzaku{} observations. We checked all the data sets for any discrepancy between them. We excluded the XIS data in the $1.7-2.3\kev$ from our spectral analysis due to calibration uncertainties. 

\begin{figure*}

   \includegraphics[width=7.5cm,height=6.8cm,angle=0]{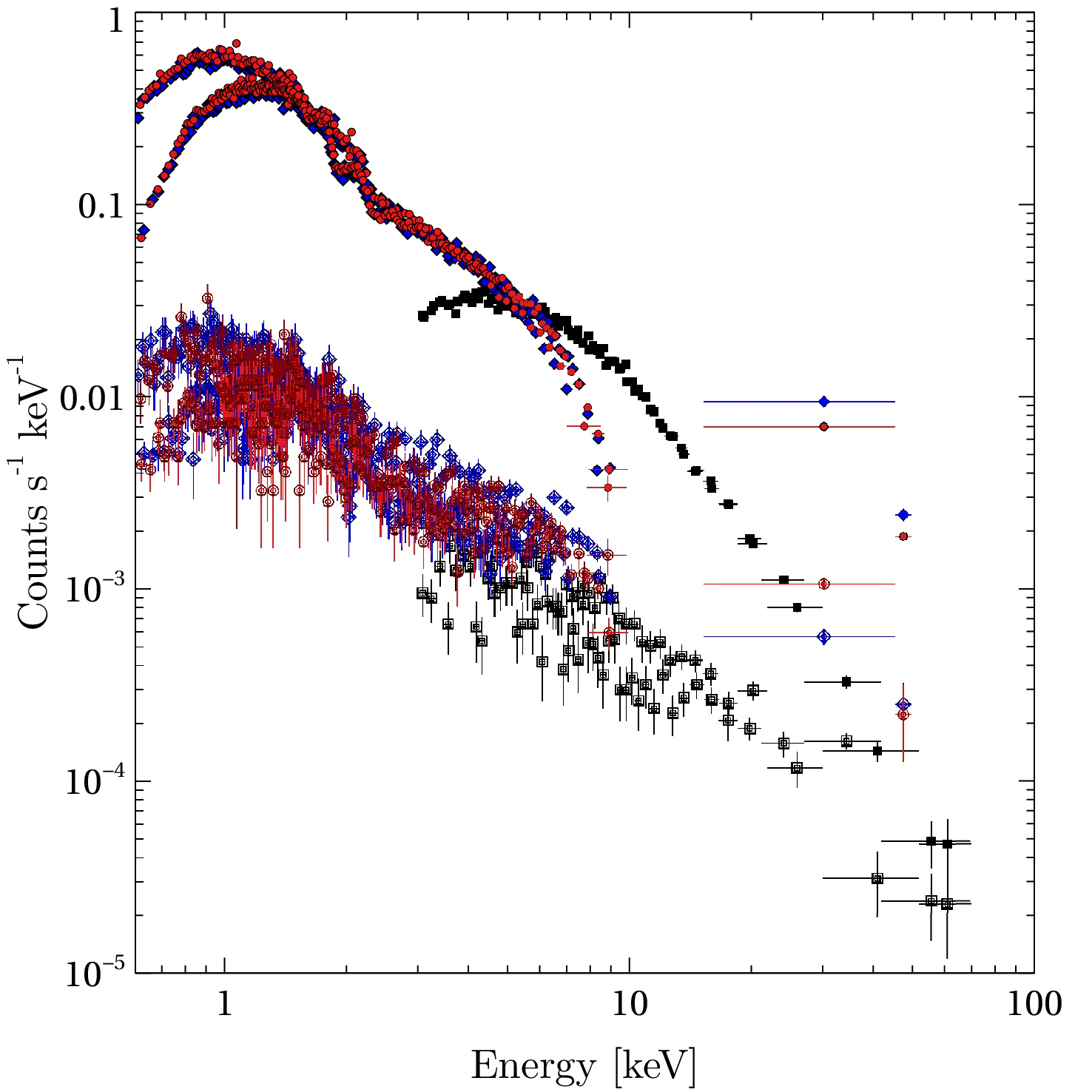}
   \includegraphics[width=7.0cm,height=7.0cm,angle=0]{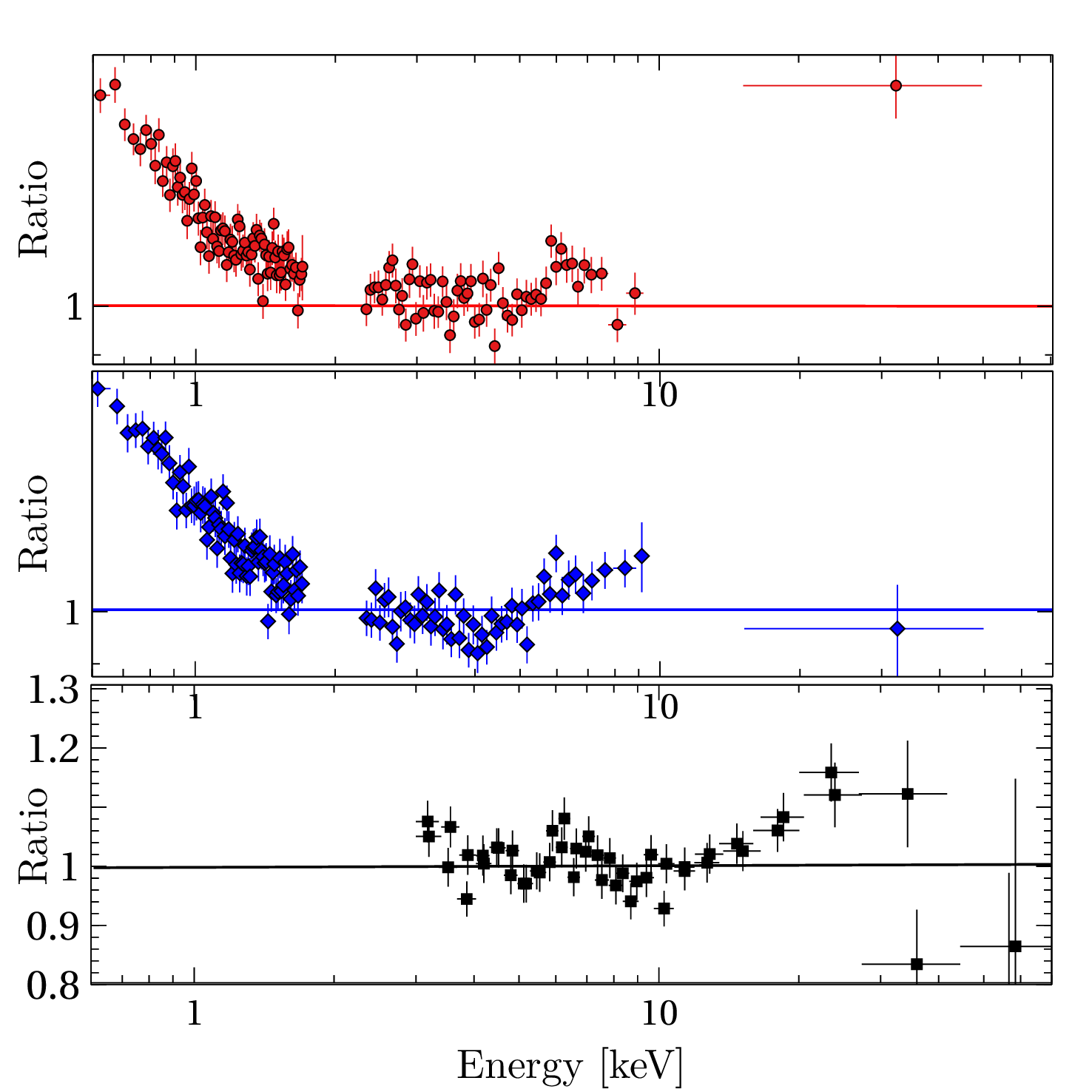}
   
   \caption{{\it Left:} Spectral data from Obs-1 (circles), Obs-2 (diamond),  and Obs-3 (squares).  Filled symbols represent the source and the open symbols represent corresponding background spectra. {\it Right:} The ratio of spectral data and  the  best-fitting $2-5\kev$ absorbed powerlaw model for Obs-1 (top), Obs-2 (middle) and Obs-3 (bottom). The XIS data sets are rebinned in Xspec for clarity.
  \label{residuals_absorbed_powerlaw}}
  
\end{figure*}

We started our spectral analysis by fitting a simple powerlaw model modified by the Galactic absorption ($N_H = 12.7\times10^{20}\cmsqi$; \citealt{2005A&A...440..775K}) to the  XIS0+XIS3, the XIS1, and the HXD/PIN data jointly. We used the absorption model {\tt tbabs} with the Verner cross sections and Wilms abundances. We used an energy-independent multiplicative factor to account for the relative normalizations of different instruments. We fixed the multiplicative factor to 1 for the XIS0+XIS3  data, and  varied for the XIS1 data. The normalization of the HXD/PIN data relative to the XIS0 data was fixed to $1.18$ for Obs-1 (HXD nominal position) and to $1.16$ for Obs-2 (XIS nominal position). First, we fitted the absorbed powerlaw model to the $2-5\kev$ band, and then we extrapolate the powerlaw model down to $0.6\kev$, and also include the $15-50\kev$ high energy PIN data. We found a prominent excess of counts below $2\kev$, a weak hard excess beyond $10 \kev$, and a hint of Fe emission line in both the \suzaku{} observations. We followed a similar approach for the \nustar{} data and fitted the $3-5\kev$ band with the absorbed powerlaw model, and then extrapolated  to $70\kev$. We again found hints for a broad Fe emission line and a hump indicating the presence of X-ray reflection features in the \nustar{} spectral data. The spectral data from all the observations (Obs-1, Obs-2 and Obs-3, hereafter for the two \suzaku{} and one \nustar{} observation, respectively) and the ratios of the observed spectral data and the model are plotted in Figure~\ref{residuals_absorbed_powerlaw}.

 \suzaku{}, \nustar{} and \swift{} provide us with the broadband UV to X-ray spectral energy distribution. To analyse the multi-epoch broadband spectra,  we organised the spectral analysis into three parts: above 3$\kev$, full X-ray band, and joint X-ray UV. This will enable us to study  the reflection feature in the datasets and also to disentangle the contribution from disc/corona and jet. Analysis of individual observations is followed by simultaneous spectral analysis of broadband spectra.

\subsection{Above 3$\kev$}
\subsubsection{The \suzaku{} data}

 We begin our analysis with the two \suzaku{} observations. In order to investigate the excess emission around $6\kev$, we first introduce a narrow Gaussian emission line ($\sigma = 0.01 \kev$) to the best-fit absorbed powerlaw model obtained for the data above $3\kev$ ($\cd=442/411$ and $297/275$ for Obs-1 and 0bs-2, respectively). This model slightly improved the fit for Obs-2 ($\dc \sim 15$ for two additional parameters) and the fit remained similar for Obs-1 ($\dc \sim 3$). Next, we replaced the narrow Gaussian with a broad line and it improved the fit significantly by $\dc \sim 14$ and $\dc \sim 13$ over the narrow Gaussian line model for Obs-1 and Obs-2, respectively with no additional parameters. The iron line energies are found to be $6.32\kev$ and $6.47\kev$ for Obs-1 and Obs-2, respectively, but we are unable to constrain the width of emission line and have assumed a broad line width of $0.1\kev$. 

Alternative models like a partial covering absorber model (neutral or ionized) may result in a similar shape of broadband X-ray spectrum. The partial covering (pc) model with neutral absorption ($zpcfabs$) multiplied with the absorbed (Galactic) powerlaw model resulted in $\cd=445/409$ for Obs-1 and $\cd=280/273$ for Obs-2, respectively. 
We  also applied the partial covering ionised absorption model by replacing the ($zpcfabs$) model with $zxipcf$ and performed fitting to the three data sets. This resulted in poor statistics with $\cd=443/408$ (Obs-1) compared to broad Gaussian line model ($\cd=425/409$). For Obs-2 these values are $\cd=278/272$ ($zxipcf$ model) and $\cd=269/273$ ($Gaussian$ line model), respectively.

 We also tested the possible presence of a relativistically broadened iron emission line by replacing the Gaussian line  with the {\tt RELLINE}~\citep{2010MNRAS.409.1534D} model. This model allows flexibility in the emissivity of the accretion disk and assumes a strongly curved space around a spinning black hole. We fixed the Fe line rest frame energy to the Gaussian line centroid and found a similar best-fit with $\chi^{2}/dof=423/405$ and $272/269$ for Obs-1 and Obs-2, respectively. Again, we were not been able to constrain the line width or the spin parameter. These results indicate the presence of a possible broad Fe emission line in the spectra of 1H~0323+342. The set of models used to fit this excess around $6\kev$ and their respective $\cd$ are quoted in Table~\ref{model_compare} for a better comparison between them.


To further characterise the broad Fe emission line and the associated reflection continuum, we used the blurred reflection model {\tt relxill} to fit our three data sets. The {\tt relxill}~\citep{2014ApJ...782...76G} is a physically motivated, angle-dependent disc reflection model which describes the origin of $Fe K_{\alpha}$ line along with reflection continuum. This model includes both the {\tt XILLVER} reflection code ~\citep{2010ApJ...718..695G} and the {\tt RELLINE} code which takes into account all the xillver-reflection spectra that occur at each point of the accretion disk and calculates proper emission angle of radiation. This {\tt XILLVER} model also takes into account the radiative transfer, the energy balance, and the ionisation equilibrium in a Compton-thick, plane-parallel medium to explain the overall spectrum. We are unable to constrain the inner radius or the spin of the black hole. The {\tt relxill} best-fit parameters are found to remain similar within  errors for the two \suzaku{} datasets (see Table~\ref{relxill fit}). The blurred reflection model with black hole spin parameter fixed at the maximal value ($a \sim 0.99$) provided statistically good fit for both  datasets. The reflection fraction is found to be $0.47^{+0.56}_{-0.58}$ and $0.67^{+1.08}_{-0.23}$ for Obs-1 and Obs-2, respectively.

\subsubsection{The \nustar{} data}

The absorbed powerlaw model here provides a satisfactory fit with $\chi^{2}=649$ for 621 dof. The inclusion of a narrow Gaussian line resulted in a marginal change in the fit  ($\dc \sim 3$ for two additional parameters). A broad Gaussian line instead of the narrow line further improved the fit by $\dc \sim 5$ to $\cd=641/618$. The best-fit Gaussian line resulted in an upper limit of $0.27\kev$ for the line width, and the line energy is found to be $6.45\kev$. Replacing the Gaussian line with the  {\tt RELLINE} resulted in a similar best-fit of $\cd=642/618$, and supports the broad nature of the Fe emission line. The {\tt relxill} model further improved the fit to $\cd=625/618$. We were unable to constrain the key parameters such as the inner radius or the spin of the black hole. The reflection fraction was found to be $0.49^{+0.16}_{-0.22}$. The best-fit parameters are consistent with the \suzaku{} observations (see Table~\ref{relxill fit}).

\subsubsection{The joint spectral analysis}

 In order to improve the errors on the best-fit parameters, we performed a joint spectral analysis of two \suzaku{} and one \nustar{} spectral datasets above $3\kev$. Apart from the powerlaw photon index and norm all other parameters were tied for the three datasets. We tested the models fitted to the individual datasets. First, we used the absorbed powerlaw model with a narrow Gaussian, and next replaced the narrow emission line with a broad one (the model reads as $po+bb+zgauss$).  The best-fit models resulted in $\cd=1338/1187$ (narrow Gaussian), 1328/1187 (broad Gaussian), respectively. The relline model instead of broad Gaussian provided a similar fit with $\cd=1328/1187$. 
In the case of the {\tt relxill} model, we tied the spin parameter $a$, the inner radius $r_{in}$, the emissivity index $\beta$ and the inclination angle $i$ for the three datasets. The {\tt relxill} model  provided a significant improvement over the simple absorbed powerlaw fit ($\cd= 1398/1197$) and resulted in a  good fit with $\chi^{2}/dof=1276/1190$ with spin parameter $a>0.90$ and inner radius $r_{in}<4.47r_{g}$, respectively (See Table~\ref{relxill fit}). For a RLNLS1, we are looking down a jet, and therefore the viewing angle of the spin-axis is small but we were unable to constrain the inclination angle and fixed it to the best-fit value (38.9 degrees) obtained from the broadband data. We also fixed the inclination to 10 degrees, and found that it produces a slightly poor fit with $\cd= 1277/1190$ but more importantly, this fit did not affect other parameters significantly and remained within errors. So, the value of inclination angle does not significantly affect the overall best-fit statistics at least for the above $3\kev$ fit. These results clearly show that {\tt relxill} model is able to fit the reflection features for all three datasets and constrain the parameters. We used this model below to fit the full X-ray band that includes the prominent soft excess below $3\kev$.

 \begin{table*}
 \centering
 {\large
   \caption{The comparison of different models used to fit the excess around $6\kev$ for the two \suzaku{} observations and the joint fit of 1H~0323+342.
 \label{model_compare}} 
  \begin{tabular}{|c|c|c|c|} \hline 
 Models                                  & Obs-1     & Obs-2   & Simultaneous fit \\\hline
 Used					& $\cd$	    & $\cd$   & $\cd$             \\\hline
 model 1	                        &442/411    &297/275  &1398/1197\\
 model 1 $+$ narrow Gaussian		&439/409    &282/273  &1338/1187\\	  
 model 1 $+$ zpcfabs			&445/409    &280/273  &1359/1191\\	  
 model 1 $+$ zxipcf  		        &443/408    &278/272  &1353/1190\\
 model 1 $+$ broad Gaussian		&425/409    &269/273  &1328/1187\\	  
 model 1 $+$ relline			&423/405    &272/269  &1332/1186\\	  

   \hline \hline
 \end{tabular}} \\ 
 Notes: Model 1 is above $3\kev$ powerlaw fit modified by the Galactic absorption. \\
 The normalization parameter of all the models were made free for each observations.
 \end{table*}

\subsection{The full X-ray band}

\begin{figure*}
  \includegraphics[width=8.0cm,height=7.8cm,angle=0]{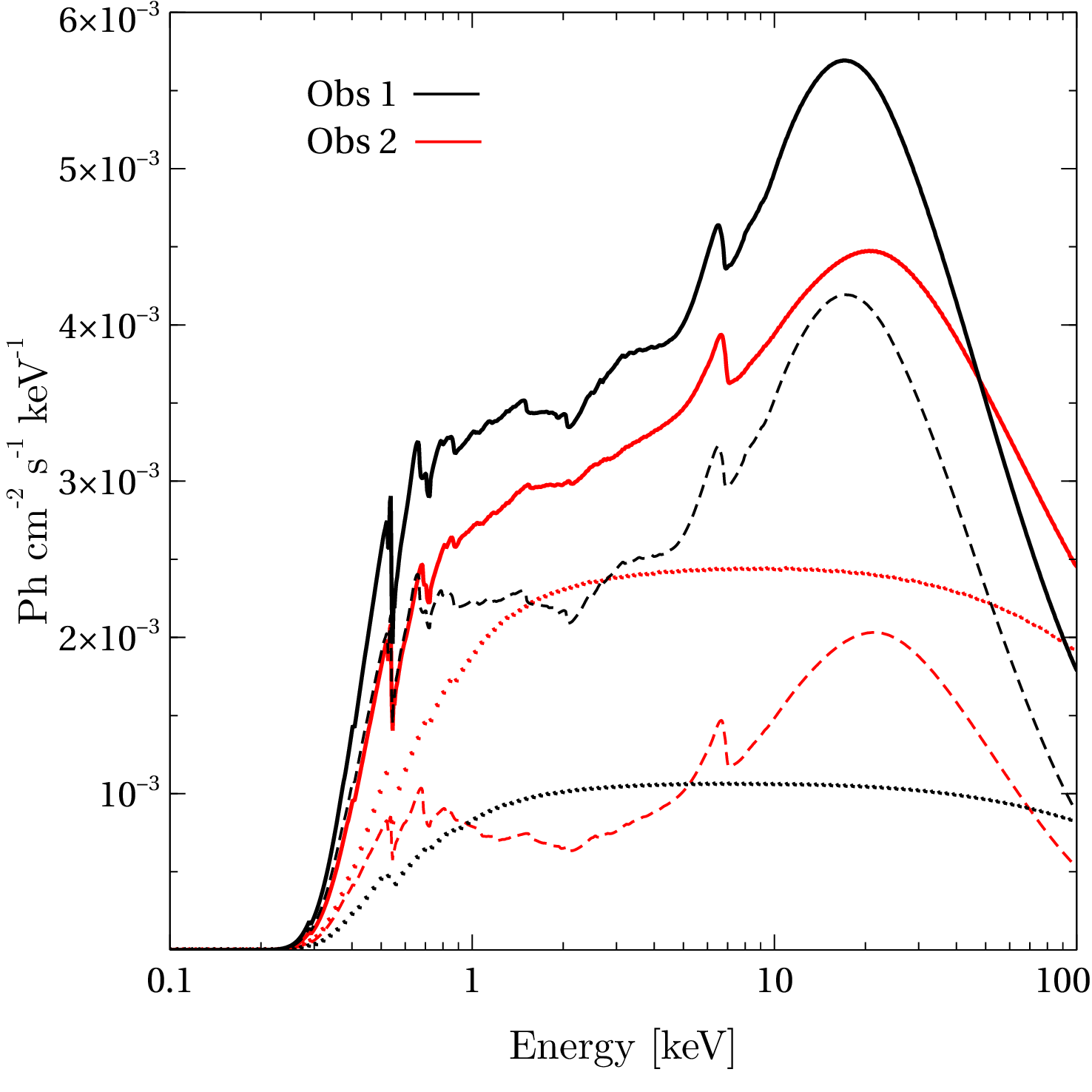}
  \includegraphics[width=8.0cm,height=7.8cm,angle=0]{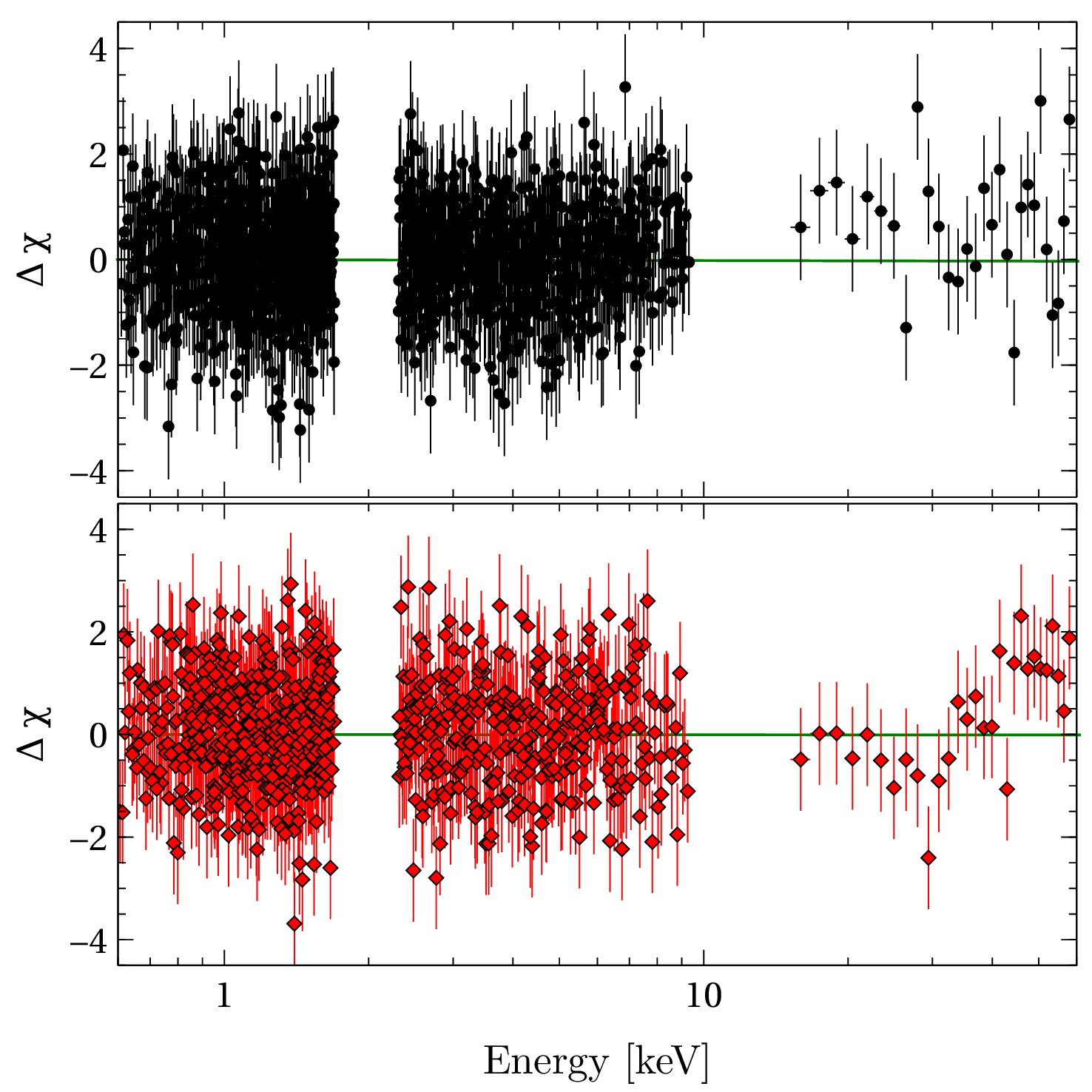}
  \caption{{\it Left:} Total absorbed {\tt relxill} best fit model (solid), reflected component (dashed), and powerlaw components (dotted) are shown seperately of XIS0+3 for Obs-1 (in Black) and Obs-2 (in Red). The reflection fraction is set to -9(Obs-1) and -3(Obs-2) to show the reflected component and to 0 to show the powerlaw components respectively. {\it Right:} We show the residuals of the respective model for Obs-1 (Top panel) and Obs-2 (Bottom panel).}
  \label{relxill_fit_plot}
  
\end{figure*}

\begin{figure*}
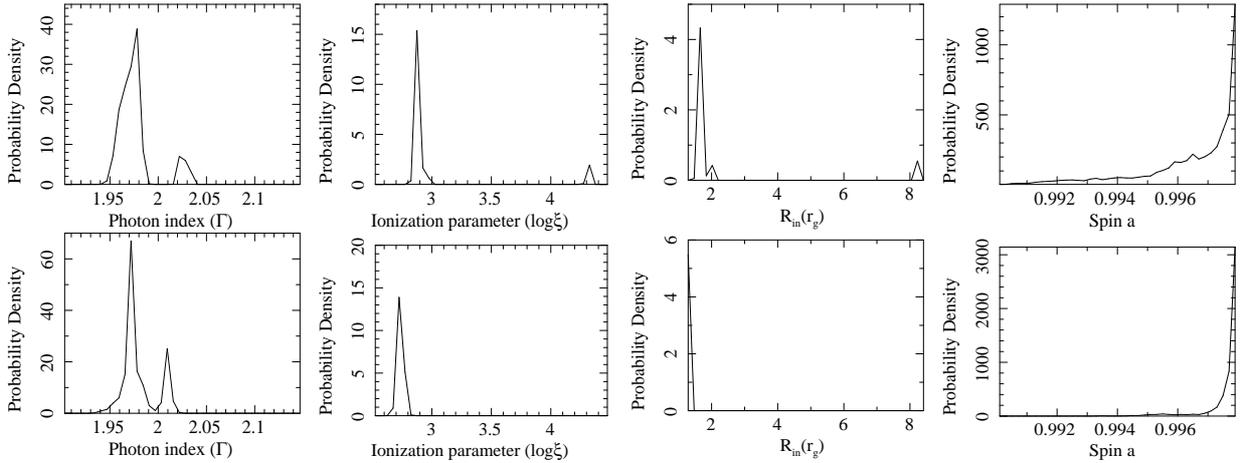

  \centering

  \includegraphics[width=3.0cm,height=4cm,angle=-90]{Probability_distribution_gamma_obs1.ps}
  \includegraphics[width=3.0cm,height=4cm,angle=-90]{Probability_distribution_logxi_obs1.ps}
  \includegraphics[width=3.0cm,height=4cm,angle=-90]{Probability_distribution_rin_obs1.ps}
  \includegraphics[width=3.0cm,height=4cm,angle=-90]{Probability_distribution_spin_obs1.ps}

  \includegraphics[width=3.0cm,height=4cm,angle=-90]{Probability_distribution_gamma_obs2.ps}
  \includegraphics[width=3.0cm,height=4cm,angle=-90]{Probability_distribution_logxi_obs2.ps}
  \includegraphics[width=3.0cm,height=4cm,angle=-90]{Probability_distribution_rin_obs2.ps}
  \includegraphics[width=3.0cm,height=4cm,angle=-90]{Probability_distribution_spin_obs2.ps}

\caption{Probability distribution of $\Gamma$, Ionization parameter $log \xi$, Inner radius $R_{in}$ and spin $a$ for Obs-1({\it Top row:}) and for Obs-2 ({\it Bottom row:}) for the {\tt relxill} model employed.}
\label{probability_relxill}
  
\end{figure*}

Next we use the full X-ray band including the soft excess, the Fe emission line and hard X-ray emission. 
An absorbed powerlaw model provided a poor fit ($\chi^{2}/dof=1968/789$ for Obs-1 and $\cd=2811/1100$ for Obs-2) mainly due to the presence of soft X-ray excess below $2\kev$. The inclusion of a simple {\tt bbody} model for the soft excess improved the fit with $kT\sim 0.14 \kev$. The photon index $\Gamma$ is found to be $\sim 1.90$ for both \suzaku{} observations. Again, a hint of residual around $6\kev$ is observed. Usually, both phenomenological and physical models are used to fit the soft excess. Previously the Obs-1 was fitted with disc reflection models ~\citep{2013MNRAS.428.2901W,2014ApJ...789..143P,2015AJ....150...23Y} and produced a good fit. 
We also used the blurred reflection model {\tt relxill} to the full X-ray band. This model provided a good fit for the both \suzaku{} observations with $\chi^{2}$/dof=868/784 and $\chi^{2}$/dof=1172/1091. The addition of the {\tt pexmon} model did not improve the fit and rules out any significant contribution from of any distant neutral reflection. We note that the inclination angle obtained for Obs-2 is unphysically high ($i= 80\pm4$ degrees) for a NLSy1 galaxy consequently we avoid any further reference to this value and present the results obtained with the inclination fixed at Obs-1 best-fit value ($\sim39$ degrees). We were still unable to constrain some of the reflection parameters of interest, e.g. the reflection fraction (R) for Obs-1 and the emissivity index {\it q} for both \suzaku{} observations. The iron abundance($\sim 0.83$) is consistent with the Solar value for Obs-1 but is slightly higher($\sim 3$) for Obs-2. Other parameters are $log \xi \sim 2.8$,  $a \sim 0.99$ and $R_{in}(r_{g}) \sim 1.4$ (See Table~\ref{relxill fit}) and are consistent with the observations. These results suggest the presence of an accretion disk that lies well within $6 r_{g}$, the last stable orbit around a non-rotating Schwarzschild  black hole corroborating the high spin state of the central black hole. These results are in agreement with that of \citealt{2013MNRAS.428.2901W} (for Obs-1), but contradicts \cite{2015AJ....150...23Y} who found a spin parameter, though poorly constrained, to be < 0.13 at the 90\% confidence level for the same observation. Here we note that \cite{2015AJ....150...23Y} used the combined FI spectrum of \suzaku{} and only $0.5-10 \kev$ energy band was used. Generally, broad-band X-ray spectrum which includes the reflection hump in the $20-50 \kev$ can provide a reliable and stronger constraint on the black hole spin parameter and makes our results more definitive. In Fig.~\ref{relxill_fit_plot} we have shown the two data sets, the best-fitting models and the respective residuals for the {\tt relxill} model. We have performed a Monte Carlo Markov Chain (MCMC) analysis for selected best-fit parameters to make sure that they are not stuck in any local minima. The fit has produced a similar probability density (See Fig.~\ref{probability_relxill}).

Next, we included the \nustar{} broadband data along with the two \suzaku{} datasets and performed joint analysis to investigate the reflection features in the broadband X-ray spectra. 
A significant improvement over the absorbed powerlaw is noticed when we replaced the powerlaw by {\tt relxill} model. However, the reflection fraction and the black hole spin are higher compared to that inferred from the data above $3\kev$ (see Table~\ref{relxill fit}). The significantly different reflection parameters suggest that the entire soft X-ray excess may not be arising from the blurred reflection alone. Hence, we explored alternative models for the soft X-ray excess.


\begin{table*}
\footnotesize
\centering
  \caption{The best-fit parameters for two \suzaku{} and \nustar{} observations of 1H~0323+342 for the absorbed {\tt relxill} model.\label{relxill fit}}
{\renewcommand{\arraystretch}{1.3}
\setlength{\tabcolsep}{4pt}

  \begin{tabular}{lllllllll} \hline
Component  & parameter                & ~~~~~~~Obs-1 &              &~~~~~~~Obs-2  &               &Obs-3      & ~~~~~~~~Joint fit & \\ \hline
           &                          & Broadband & Above $3\kev$ &  Broadband & Above $3\kev$ &           &Broadband & Above $3\kev$ \\ \hline
Gal. abs.  & $N_{H} (1)$ & $ 12.7$(f) & $ 12.7$(f) & $ 12.7$(f) & $ 12.7$(f)  & $ 12.7$(f) & $ 12.7$(f) & $ 12.7$(f)\\ 

relxill    &  $A_{Fe}$                & $0.8^{+0.2}_{-0.2}$ & $1$(f) & $3.0^{+1.7}_{-1.2}$ & $1$(f) &$1$(f) & $0.9^{+0.3}_{-0.1}$ &$1$(f)\\ 
           &  $log\xi (2)$            & $2.9^{+0.1}_{-0.1}$ & $2.3^{+0.4}_{-0.4}$ & $2.7^{+0.1}_{-0.3}$   & $3.3^{+0.2}_{-0.3}$   &$<2.2$ & $0.4^{+0.1}_{-0.3}$ & $2.3^{+0.2}_{-0.2}$\\ 
           & $ \Gamma $               & $1.97^{+0.04}_{-0.03}$ & $1.82^{+0.08}_{-0.09}$ & $1.97^{+0.03}_{-0.02}$   & $1.83^{+0.08}_{-0.07}$   & $1.92^{+0.03}_{-0.05}$   & $2.03^{+0.02}_{-0.07}$ & $1.84^{+0.10}_{-0.09}$\\
           &  $n_{rel}(10^{-5})^a$    & $1.6^{+0.1}_{-0.1}$ & $6.2^{+0.6}_{-0.7}$ & $3.7^{+0.4}_{-0.6}$   & $3.7^{+0.3}_{-0.5}$   & $4.6^{+0.4}_{-0.3}$   & $4.8^{+0.5}_{-0.4}$ & $6.1^{+0.3}_{-0.6}$\\
           &   $ q$                   & $>9.4$  & $6$(f)  & $>9.3$  & $6$(f)& $6$(f) & $>8.4$ & $6$(f)\\
           &   $ a$                   & $>0.972$ & $0.998$(f) & $>0.991$ & $0.998$(f)  & $0.998$(f) & $>0.996$ & $>0.900$\\
           &   $R(refl frac) $        & $>8.39$  & $0.47^{+0.56}_{-0.58}$  & $3.02^{+0.69}_{-0.86}$ & $0.67^{+1.08}_{-0.23}$ & $0.49^{+0.16}_{-0.22}$ & $8.30^{+0.65}_{-2.45}$ & $0.55^{+0.43}_{-0.45}$\\
           &   $ R_{in}(r_{g})$       & $1.60^{+0.04}_{-0.11}$ & $<2.48$ & $<1.40$ & $<2.74$ & $<5.25$ & $<2.37$ & $<4.47$\\
           &   $ R_{br}(r_{g})$       & $2.99^{+0.19}_{-0.21}$ & $6$(f) & $2.68^{+0.13}_{-0.15}$   & $6$(f) &$6$ (f) & $<3.08$  & $<7.17$\\
           &   $ R_{out}(r_{g})$      & $400$ (f) & $400$ (f)  & $400$ (f) & $400$ (f)& $400$ (f) & $400$ (f) & $400$ (f)\\
           &   $i(degree) $           & $38.9^{+2.4}_{-4.2}$& $38.9$(f) & $38.9$ (f) & $38.9$ (f)& $38.9$ (f)& $38.6^{+2.1}_{-4.8}$ & $38.6$ (f)\\ 
           &  $ f_{0.6-2\kev}^{b}$    & $8.12$  & -- & $6.61$  & -- & $1.12$ & -- & --\\
           &  $ f_{2-10\kev}^{b}$     & $10.50$ & -- & $8.91$  & -- & $8.32$ &-- &--\\
           &  $ f_{0.6-10\kev}^{b}$   & $18.60$ & -- & $15.50$ & -- & $9.33$ &-- &--\\\hline
           & $\cd $                   & $1164/1091$   & $427/408$    & $865/785$    & $287/272$ & $625/618$ &$2673/2284$ &$1276/1190$ \\\hline 
\end{tabular}} \\ 
Notes: (f) indicates a frozen parameter.
(a) $n_{rel}$ reperesents normalization to relxill model component. 
(b) The unabsorbed flux in units of $10^{-12}$ $\funit$.
(1):in units of $10^{20}cm^{-2}$; (2): in units of $\xiunit$.
\end{table*}

\subsection{The joint X-ray/UV spectral analysis \& the intrinsic disc Comptonization model}

 Apart from the disc reflection model, the intrinsic disc Comptonization~\citep{2012MNRAS.420.1848D} is another physically motivated model which can explain the soft X-ray excess in a self-consistent way and can provide us useful information about the temperature and optical depth as well as the size of the optically thick, warm corona. This model enables us to study the intrinsic disc emission, peaking in the optical/UV band, in addition to the soft X-ray excess and the powerlaw component. We used the simultaneous  optical/UV data from \swift{}/UVOT and the \suzaku{} X-ray data. The \nustar{} data were not included as there are no simultaneous archival optical/UV data available. 

 The intrinsic disc Comptonization model {\tt optxagnf} assumes that 
 the gravitational energy released in the accretion disk powers all the optical/UV, the soft X-ray excess and the powerlaw emission. The optical/UV emission arises in the form of  a colour temperature corrected blackbody down only to $r_{corona}$. Inside $r_{corona}$ the energy released is divided into the warm thermalized disc and optically thin, hot corona to produce the soft excess and the powerlaw continuum, respectively. This self-consistent model imposes a change in the disc structure inside the radius $r_{corona}$. It  assumes that all the material are accreting through the outer thin disc, and uses energy conservation law to describe and constrain the luminosity of the soft excess and the power-law tail.

We used the optical/UV fluxes in six bands measured with the \swift{} to constrain the thermal emission from the disc. 
We used the  {\tt optxagnf} model modified by the Galactic X-ray absorption and optical/UV reddening and fitted the simultaneous broadband \swift{} UVOT and \suzaku{} XIS/PIN data. The Galactic extinction ($E_{V-B} = 0.21$; \citealt{1998ApJ...500..525S}) is taken into account by the {\scshape REDDEN} model. We fixed the {\tt optxagnf} normalization at unity, as the flux is set by the physical parameters of black hole mass ($\rm M_{BH}$), relative mass accretion rate ($\rm  L/L_{Edd}$), black hole spin ($a^{*}$) and luminosity distance of the source ($\rm D_L$). We used fixed values for the black hole mass ($\rm M_{BH}$) and the luminosity distance of the source ($\rm D_L$) and varied the accretion rate ($\rm L/L_{Edd}$), the $r_{corona}$ and the spin parameter $a^{*}$. $\rm M_{BH}$ is fixed at $10^{7}\msol$,~\citep{2010ApJ...717.1243G} and ($\rm D_L$) is fixed at $260 \mpc$.

Initially, we fixed black hole spin parameter to zero which produces a poor
fit for both the \suzaku{} observations with $\chi^{2}$/dof=1627/1101 and 
$\chi^{2}$/dof=1588/794 for Obs-1 and Obs-2, respectively. Varying the spin parameter  
did not improve the fit for Obs-1 but reduced the $\chi^{2}$ to
1304 for Obs-2 but still remained a statistically poor fit. Therefore we fitted the \suzaku{} data first with the
{\tt optxagnf} and extrapolated to the UV band using the \swift{} data. Both observations show 
significant excess in the UV band.  Such a possibility may arise due to the contribution from the jet in the UV band.
We add a simple powerlaw model to fit the excess and this resulted in a good fit with $\chi^{2}$/dof=1204/1098 and $\chi^{2}$/dof=942/791 for Obs-1 and Obs-2, respectively. We noticed excess emission above 20$\kev$, particularly in the second observation.  We added another powerlaw for this hard X-ray excess. The fit improved for Obs-2 only ($\chi^{2}$ by $\sim5$) but remained unchanged for Obs-1. Next, we removed both the powerlaw
models and introduced a broken powerlaw. We fixed {\tt optxagnf} $\Gamma$ to 2.0 and {\it fpl} to zero. This model produces slightly better fit with $\chi^{2}$/dof=1200/1098 and$\chi^{2}$/dof=921/791 for Obs-1 and Obs-2, respectively. The best-fit parameters are  listed in Table~\ref{optxagnf bknpo fit}. The Coronal radius is found to be (in $r_{g}$) >75, relatively large for both observations. This distance marks the transition from colour temperature
corrected blackbody emission to a Comptonized spectrum. Moreover, the Eddington ratios are $\sim$0.74 and 0.16 for Obs-1 and Obs-2 respectively. This is in agreement with \cite{2015AJ....150...23Y,2014ApJ...789..143P} who found similar results using SED modelling for Obs-1. The electron temperature of the  optically thick corona producing the soft X-ray excess is $kT_{e}> 0.37\kev$ and the optical depth $\tau$ is in the range of $7.1-13.1$. These values are in agreement with that of other RLNLSy1 galaxies (e.g., \citealt{2004MNRAS.349L...7G,2011ApJ...727...31A}). 
The data sets, the best-fitting models and the deviations of the data from the model are shown in Fig.~\ref{broadband_fit}.

\begin{table}
\footnotesize
\centering
  \caption{The best-fit parameters for \suzaku{}  and \swift{} observations of 1H~0323+342 for the combined optxagnf and broken powerlaw model.
\label{optxagnf bknpo fit}}
  
{\renewcommand{\arraystretch}{1.5}
\setlength{\tabcolsep}{1pt}
  \begin{tabular}{cccccc} \hline
 Component       &   Parameter             & Obs-1                  & Obs-2           & Simultaneous &fit   \\\hline
                 &                         &                        &                 & Obs-1       & Obs-2  \\\hline
Gal. abs.        & $N_{H}{(1)}$            & $ 12.7$ (f)            & $ 12.7$ (f)     & $ 12.7$ (f) & $ 12.7$ (f)      \\
Redden           & E(B-V)                  & $0.21$ (f)             & $ 0.21$ (f)     & $ 0.21$ (f) & $ 0.21$ (f)           \\ 	   
optxagnf         & $ M_{BH}{(2)}$          & $ 1.0$ (f)             & $ 1.0$ (f)      & $ 1.0$ (f)  & $ 1.0$ (f)  \\
                 & $d{\rm~(Mpc}) $         & $ 260$ (f)             & $ 260$ (f)      & $ 260$ (f)  & $ 260$ (f) \\
                 &  $(\frac{L}{L_{E}})$    &$0.74^{+1.04}_{-0.18}$  &$0.16^{+0.05}_{-0.05}$ &$1.94^{+0.05}_{-0.55}$ &$1.06^{+0.15}_{-0.71}$ \\ 
                 &  $ kT_{e} (\kev)$       & $ >0.62 $              & $0.48^{+0.22}_{-0.11} $ & \textbf{0.28}$^{+0.40}_{-0.05} $ &\\ 
                 &  $ \tau $               & $7.7^{+0.6}_{-0.7}$    & $11.0^{+2.1}_{-1.4}$ & \textbf{5.9}$^{+4.1}_{-1.4}$ &\\
                 &  $ r_{cor}(r_{g})$      & $>96.0$                & $>74.8$                 & \textbf{61.7}$^{+21.5}_{-7.9}$   &\\
                 &  $ a $                  & $ >0.99 $              & $ >0.77$                & \textbf{0.98}$^{+0.01}_{-0.05}$  &\\
                 &  $ f_{pl}$              & $0$(f)                 & $0$(f)          & $0$(f)      & $0$(f)\\
bknpower         &  $\Gamma_{1}$    &$2.29^{+0.01}_{-0.01}$  &$2.27^{+0.01}_{-0.02}$ &$2.14^{+0.02}_{-0.02}$ &$2.19^{+0.02}_{-0.01}$\\ 
                 &  $\Gamma_{2}$    &$1.66^{+0.04}_{-0.04}$  &$1.79^{+0.04}_{-0.04}$ &$1.80^{+0.04}_{-0.03}$ &$1.83^{+0.03}_{-0.03}$\\ 
                 &  $E_{break}$     &$1.22^{+0.05}_{-0.05}$  &$1.22^{+0.05}_{-0.05}$ &$2.32^{+0.04}_{-0.05}$ &$1.90^{+0.05}_{-0.05}$\\
                 &  $n(10^{-3})$    &$2.21^{+0.23}_{-0.28}$  &$2.63^{+0.15}_{-0.19}$ &$4.06^{+0.11}_{-0.39}$ &$3.33^{+0.19}_{-0.07}$ \\ \hline
                  &  $\cd $                 & $1200/1098$           & $921/791$             & $2501/1960$  &\\ \hline \hline
\end{tabular}} \\ 
Notes: (f) indicates a frozen parameter.\\
(1):in units of $10^{20}cm^{-2}$; (2):in units of $10^7\rm M\odot$. \\
Bold values are the ones which are constrained by all observations as they are assumed to be constant.
\end{table}

\subsubsection{The joint analysis of two sets of simultaneous \swift{} and \suzaku{} observations}

We performed a joint spectral fitting  of the two \suzaku{} and \swift{} observations. Apart from the Eddington ratio and broken powerlaw parameters, all the parameters were tied between Obs-1 and Obs-2. The best-fit parameters are similar to those found for individual observations (see Table~\ref{optxagnf bknpo fit}). The Eddington ratios are found to be super Eddington, $\sim$1.94 and 1.06 for Obs-1 and Obs-2, respectively. The Coronal radius again is found to be relatively large ($61.7^{+21.5}_{-7.9} r_{g}$). The electron temperature, $kT_{e} = 0.28^{+0.40}_{-0.05}\kev$  and the optical depth, $ \tau = 5.92^{+5.15}_{-0.19}$ for the soft Comptonising component. These parameters are consistent with those obtained  for individual observations as well as for other RLNLS1 galaxies.

\subsubsection{Combined blurred reflection and intrinsic disc
Comptonization model}

\begin{figure*}

  \includegraphics[width=5.5cm,height=5.5cm,angle=0]{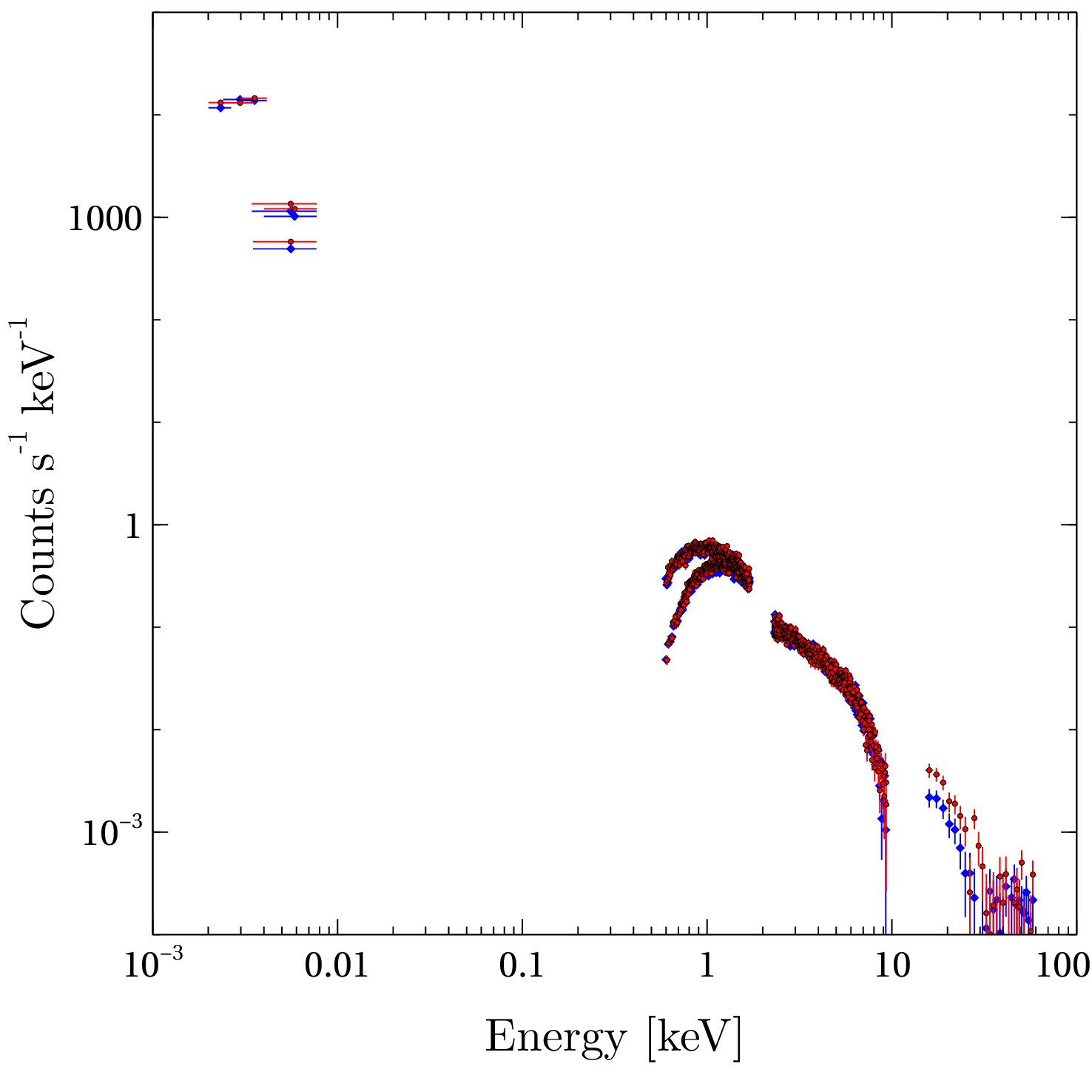}
  \includegraphics[width=5.5cm,height=5.5cm,angle=0]{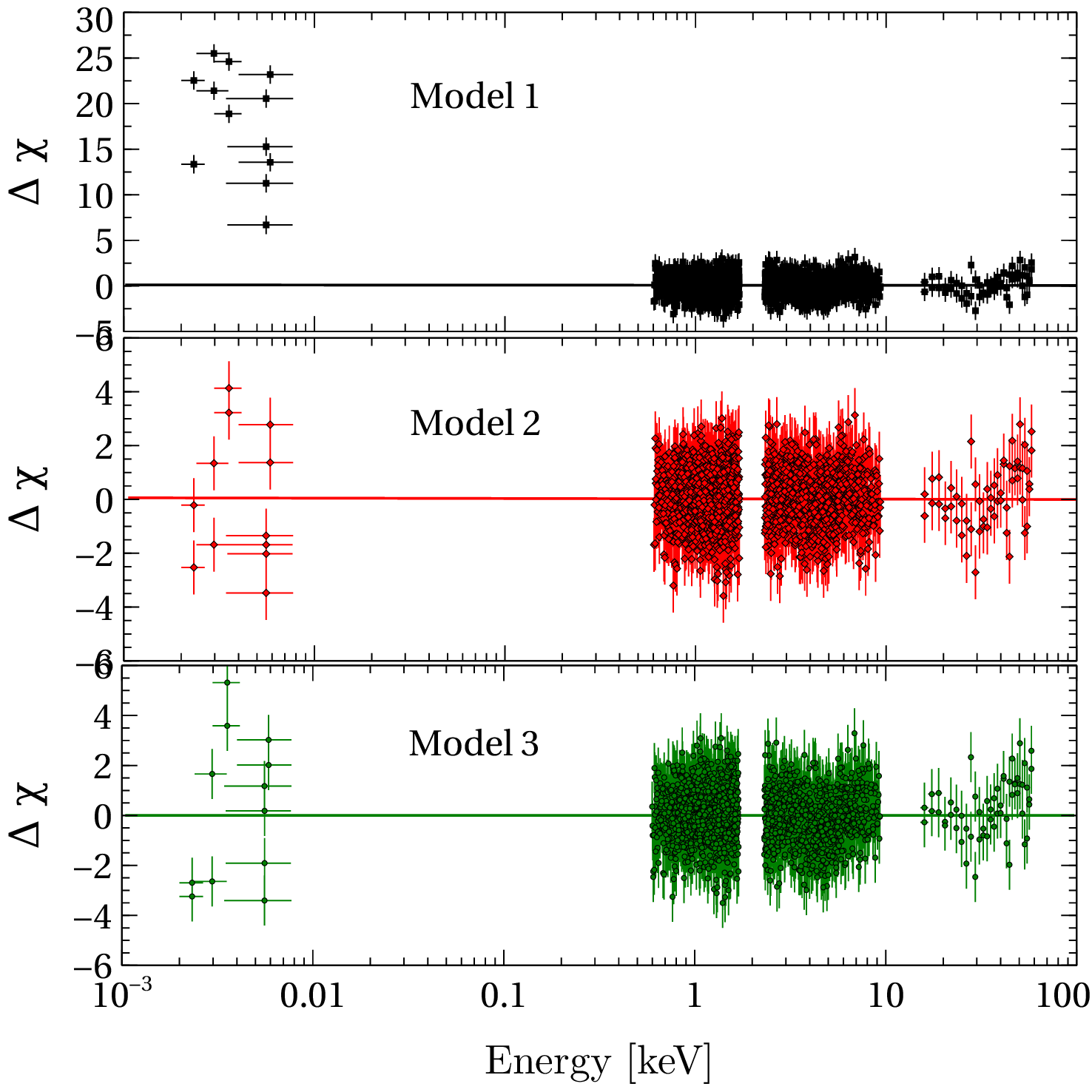}
  \includegraphics[width=5.5cm,height=5.5cm,angle=0]{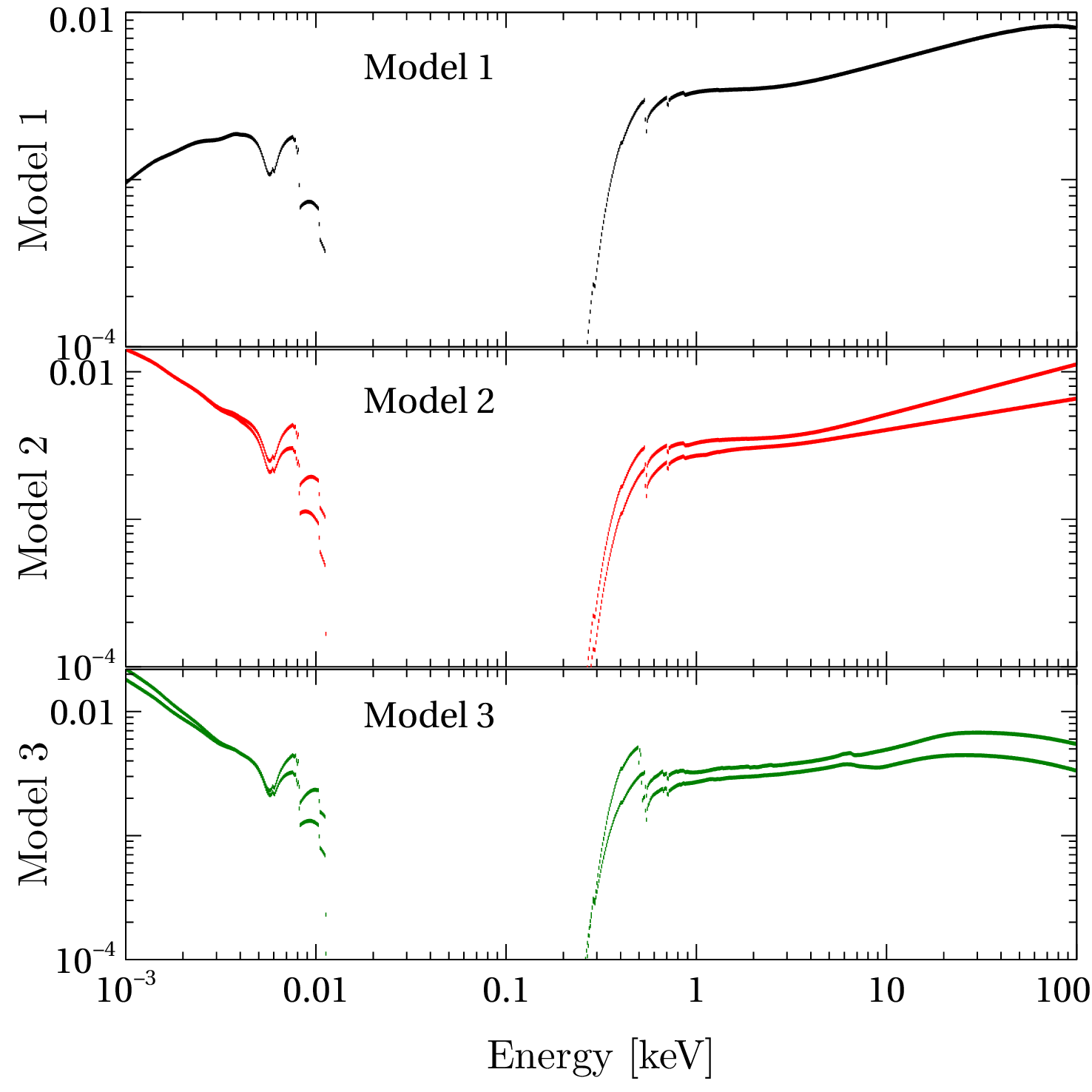}
  \caption{The figure on the left shows the simultaneous optical-UV to hard X-ray data for Obs-1(Red) and Obs-2(Blue). The middle one shows the respective residuals for the three models used to fit the broadband data in three different panels. The plot on the right shows the three total best-fit models for XIS0+3 in three different panels. The three models used are: 
Model 1 : constant$\times$tbabs$\times$redden$\times$optxagnf;
Model 2 : constant$\times$tbabs$\times$redden(optxagnf$+$bknpo);
Model 3 : constant$\times$tbabs$\times$redden$\times$(po$+$optxagnf$+$relxill).}
  \label{broadband_fit}
  
\end{figure*}

As inferred earlier based on the analysis of the data above $3\kev$, the blurred reflection is likely to have some contribution to the soft X-ray excess. Hence we fitted a complex model consisting of both the blurred reflection and the intrinsic disk Comptonisation. 
We first estimated the
contribution of the reflection to the soft excess emission by fitting the {\tt relxill} model jointly to the Obs-1 and Obs-2 above $3\kev$. The best-fit parameters are similar to those listed in Table~\ref{relxill fit}. We further extrapolate down to $0.6\kev$
and noted that the best-fit {\tt relxill} model is unable to fit the soft X-ray excess even when the parameters are allowed to vary within the 90\% confidence range. This fit resulted in a poor fit with $\chi^{2}$/dof=1443/1098 and $\chi^{2}$/dof=1116/790 for Obs-1 and Obs-2, respectively. This result further suggests that the origin of the soft excess may not be entirely due to the blurred reflection
from an ionised accretion disk. We then included the {\tt optxagnf} model for the soft excess and fitted the joint reflection and intrinsic disk Comptonisation model to the full X-ray band.  This joint model provided  a satisfactory fit. 
Next, we included the \swift{} UVOT data acquired simultaneously with Obs-1 and Obs-2, and performed the fitting. This fit again revealed a significant excess in the UV band. A steep powerlaw component is included to model the possible jet contribution in the UV band which resulted in a satisfactory fit to the broadband UV/X-ray data. The  best-fit model reads as constant$\times$tbabs$\times$redden$\times$(powerlaw$+$optxagnf$+$relxill). Here we note that optxagnf model assumes the disk to be no longer capable of producing reflection inside $r_{corona}$ and this radius is larger than the radius where the reflection is produced. Hence we quote the results with $r_{corona}$ fixed at $6 r_{g}$ in this joint model. We also found that the inclusion of the {\tt zxipcf} model with $N_{h}=1.7\times10^{21}$ and $log{\xi= -0.38}$ to the Obs-2 slightly improves the fit with a reduction in $\chi^{2}$ by 10. The best-fit parameters are quoted in Table~\ref{Combined_fit_par}. The data sets, the best-fitting models and the deviations of the data from the model are shown in Fig.~\ref{broadband_fit}.


We followed a similar approach and performed a simultaneous fit to the optical/UV to  X-ray datasets from Obs-1 and Obs-2. The results are presented with $r_{corona}$ fixed at $6 r_{g}$. Here, the {\tt relxill} parameters are fixed to above $3\kev$ best-fit for individual observations. The Eddington ratio and powerlaw parameters were free to vary, and the rest of the {\tt optxagnf} model parameters were tied between Obs-1 and Obs-2. The best-fit parameters are similar to those found for the individual observations and are listed in Table~\ref{Combined_fit_par}. 

\begin{table}
\centering
\caption{Results of spectral fits to the simultaneous \swift{} UVOT and the \suzaku{} data sets.}
\label{Combined_fit_par}
{\renewcommand{\arraystretch}{1.5}
\setlength{\tabcolsep}{1pt}
\begin{tabular}{cccccc} \hline
\hline
 Component & Parameter              & Obs 1       & obs 2        & Simultaneous & fit    \\ 
           &                        &             &              &   Obs-1      & Obs-2  \\\hline \hline
Gal. abs.  & $N_{H}{(1)}$           & $12.7$(f)   & $12.7$(f)    & $12.7$(f)    & $12.7$(f)\\
Redden     & E(B-V)                 & $0.21$(f)   & $0.21$(f)    & $0.21$(f)    & $0.21$(f)\\ 

powerlaw   & $\Gamma $              & $2.69^{+0.19}_{-0.18}$  & $2.41^{+0.22}_{-0.09}$  & $2.62^{+0.12}_{-0.08}$ & $2.53^{+0.02}_{-0.03}$\\
           & $n_{pl}(10^{-4})$      & $1.8^{+0.6}_{-0.1}$     & $10.2^{+0.8}_{-0.9}$    & $3.9^{+0.3}_{-0.4}$ & $6.8^{+0.2}_{-0.8}$\\

optxagnf   & $ M_{BH}{(2)}$         &$1.0$(f)    &$1.0$(f)       &$1.0$(f)      &$1.0$(f)\\
           & $d{\rm~(Mpc}) $        &$260$(f)    &$260$(f)       &$260$(f)      &$260$(f)\\
           &  $(\frac{L}{L_{E}})$   &$1.95^{+0.09}_{-0.70}$   &$1.69^{+0.18}_{-0.83}$  &$1.25^{+0.11}_{-0.75}$   &$0.75^{+0.38}_{-0.23}$\\ 
           &  $ kT_{e} (\kev)$      &$0.38^{+0.02}_{-0.02}$   &$0.56^{+0.01}_{-0.01}$  &\textbf{0.31}$^{+0.02}_{-0.01}$  &\\ 
           &  $ \tau $              &$>11.9$                  &$4.2^{+6.8}_{-1.4}$     &\textbf{8.7}$^{+9.0}_{-0.7}$  &\\
           &  $ r_{cor}(r_{g})$     &$ 6 $(f)    &$ 6 $(f)       & $6$(f)       &$6$(f)\\
           &  $ a $                 &$0.99 $(f)  &$ 0.99$(f)     &$ 0.99$(f)    &$0.99$(f)\\
           &  $ f_{pl}$             &$0 $ (f)    &$ 0$ (f)       &$ 0$ (f)      &$ 0$ (f)\\

relxill    &  $A_{Fe}$              & $1.0$(f)   & $1.0$(f)      & $1.0$(f)     & $1.0$(f)\\ 
           &  $log\xi{(3)}$         & $2.3$(f)   & $3.3$(f)      & $2.3$(f)     & $3.3$(f)\\ 
           & $ \Gamma $             & $1.82$(f)  & $1.83$(f)     & $1.82$(f)    & $1.83$(f)\\
           &  $n_{rel}(10^{-5})$    & $6.2$(f)   & $3.7$(f)      & $6.2$(f)     & $3.7$(f)\\
           &   $ q $                & $6$(f)     & $6 $(f)       & $6 $(f)      & $6 $(f)  \\ 
           &   $ a $                & $0.99$(f)  & $0.99$(f)     & $0.99$(f)    & $0.99$(f)      \\
           &   $R(refl frac) $      & $0.47$(f)  & $0.67$(f)     & $0.47$(f)    & $0.67$(f)\\
           &   $ R_{in}(r_{g})$     & $6$(f)     & $6$(f)        & $6$(f)       & $6$(f)\\
           &   $ R_{br}(r_{g})$     & $6$(f)     & $6$(f)        & $6$(f)       & $6$(f)\\
           &   $ R_{out}(r_{g})$    & $400$(f)   & $400$(f)      & $400$(f)     & $400$(f)    \\
           &   $i(degree) $         & $38.6$(f)  & $38.6$(f)     & $38.6$(f)    & $38.6$(f)       \\
\hline
           &  $\cd $                & $1345/1103$ & $989/795$    & $2551/1965$     &\\ \hline
\end{tabular}}

Note: (f) indicates a frozen parameter.
(1):in units of $10^{20}cm^{-2}$; (2):in units of $10^7\rm M\odot$; (3): in units of $\xiunit$.\\
Bold values are the ones which are constrained by all observations as they are assumed to be constant.
\end{table}

\begin{figure}
\includegraphics[width=8.5cm,height=4cm]{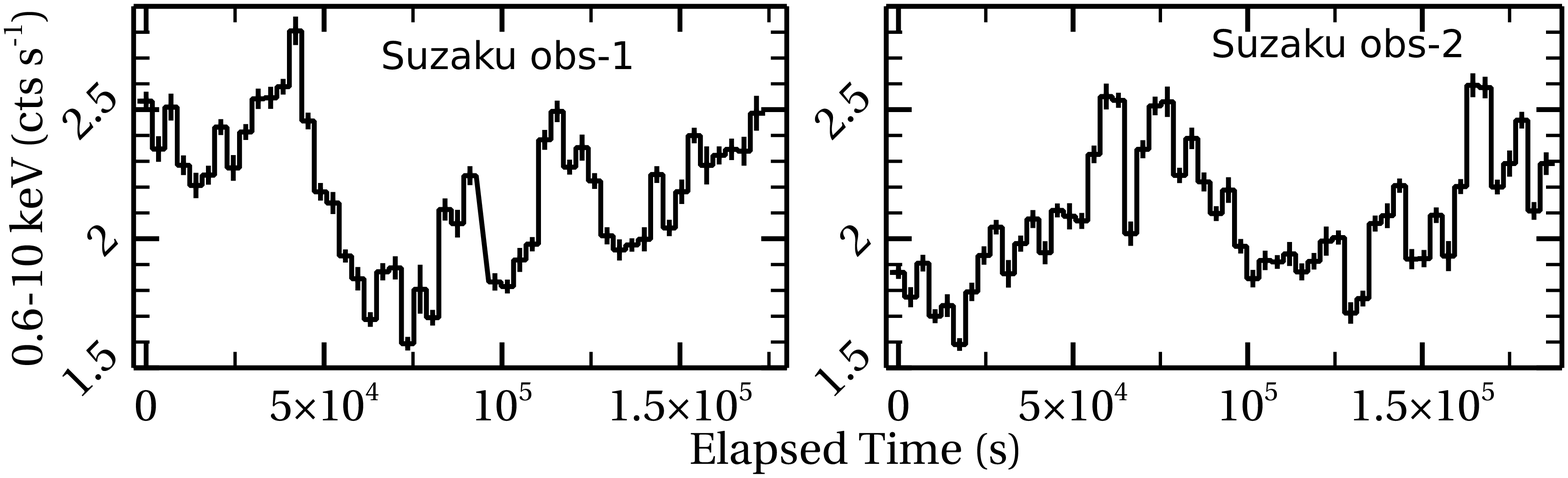}

\includegraphics[width=8.5cm,height=4cm]{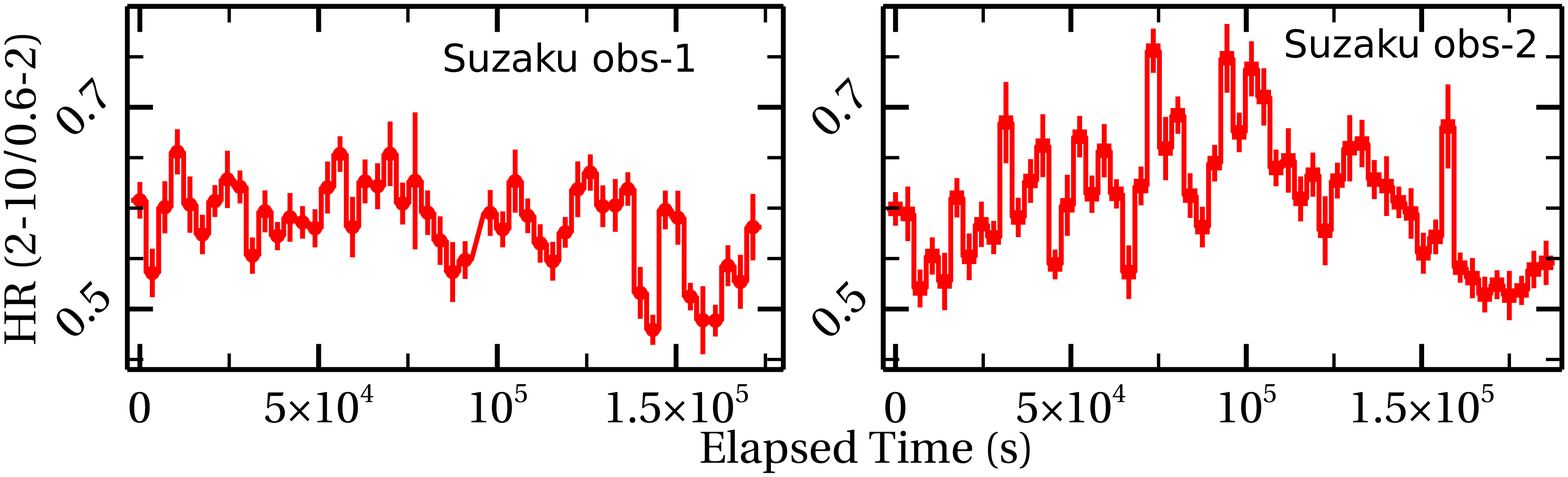}
\caption{\suzaku{}/XIS ($0.6-10\kev$) background subtracted light curves and the \suzaku{}/XIS hardness ratio (HR=$2-10\kev$/$0.6-2\kev$) of 1H~0323+342 with $3500\s$ time bins.}
\label{lc}
\end{figure}

\begin{figure*}
\includegraphics[width=5.5cm,height=4cm]{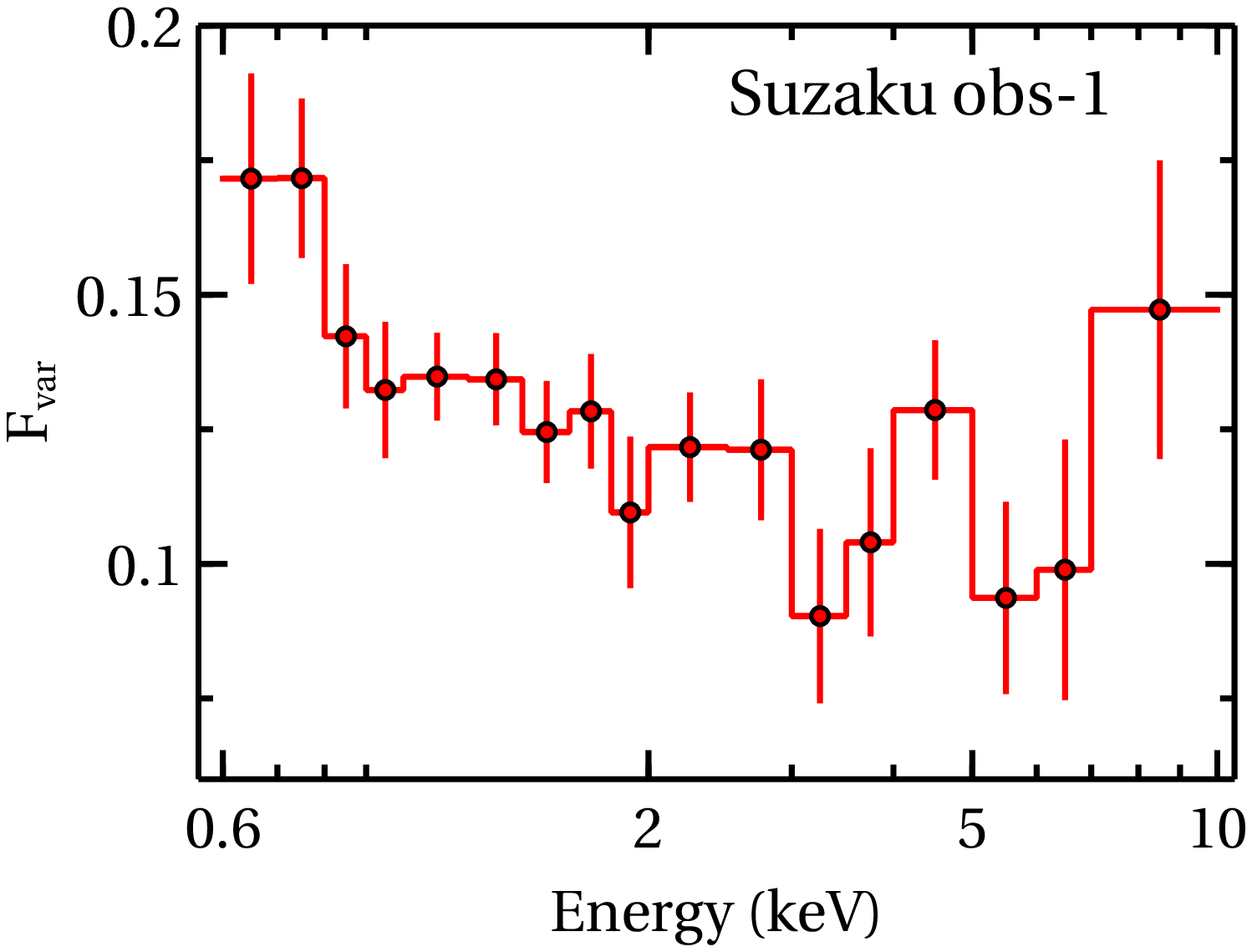}
\includegraphics[width=5.5cm,height=4cm]{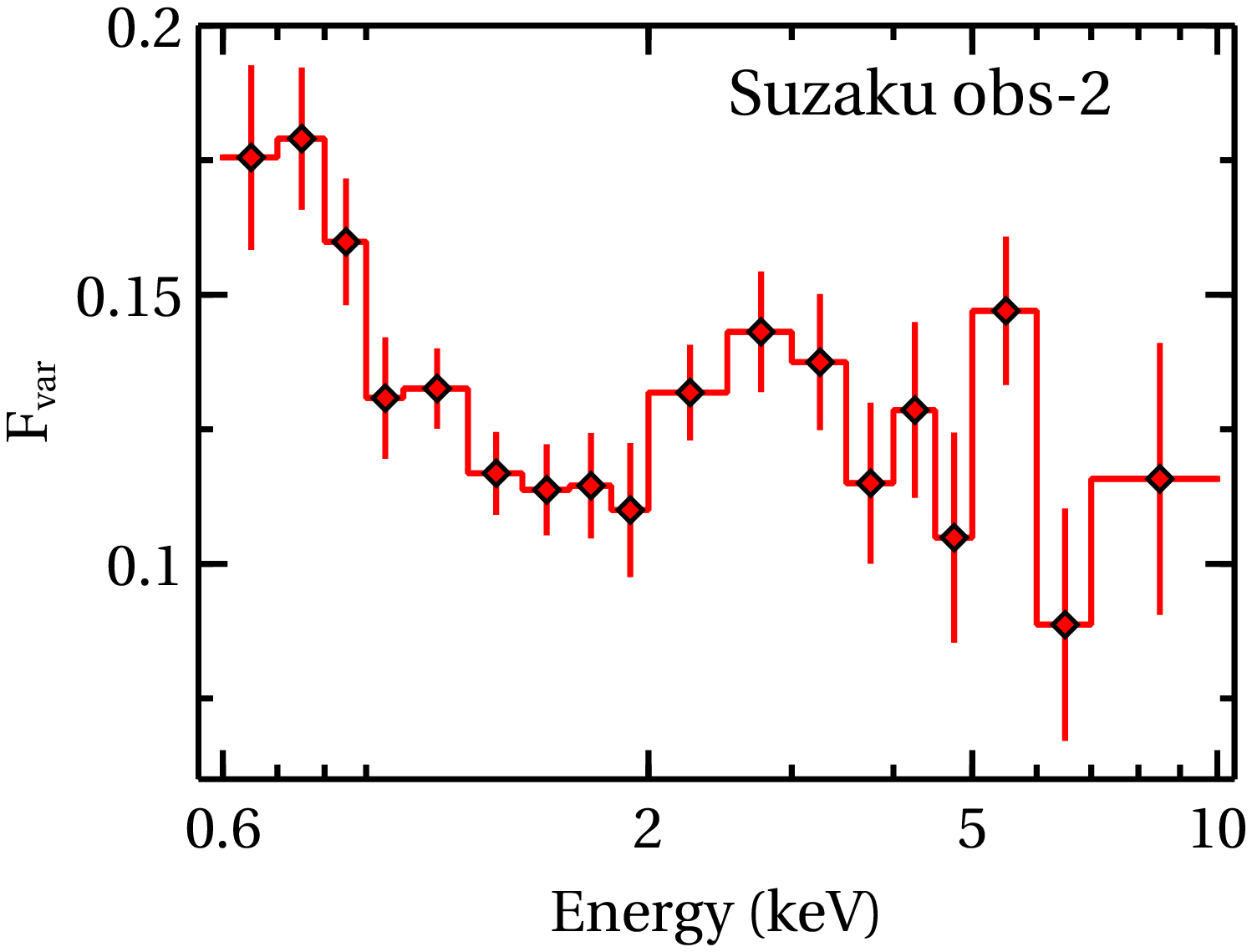}
\includegraphics[width=5.5cm,height=4cm]{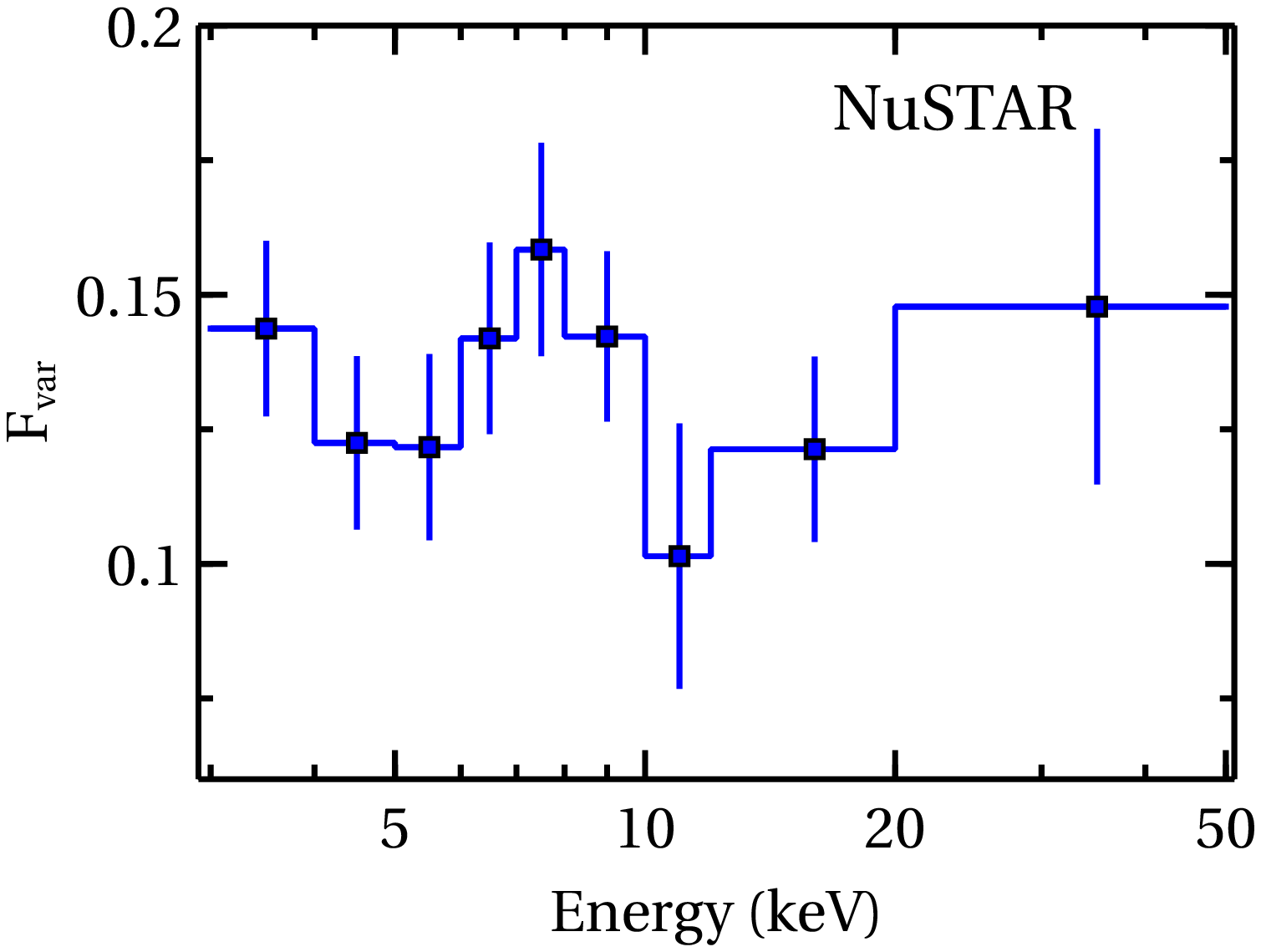}
\caption{The $0.6-10\kev$ fractional rms spectra of 1H~0323+342 derived from the \suzaku{} XIS background subtracted light curves of $171.5\ks$, $185.5\ks$ exposures for Obs-1 (left), Obs-2 (middle) and the $3-50\kev$ fractional rms spectra of 1H~0323+342 derived from the \nustar{} (FPMA+FPMB) background subtracted light curves of $195\ks$ exposure for Obs-3 (right) respectively. The time bin size is $3500\s$.}
\label{rms}
\end{figure*}

\section{X-ray Variability Analysis}

To understand the emission mechanism in a model-independent way, we study the variability properties of 1H~0323+342. First, we generate the background subtracted \suzaku{}/XIS light curve in the $0.6-10\kev$ band with time bin size of $3500\s$ which is shown in Figure~\ref{lc}. The $\chi^{2}$ test reveals significant variability in the source count rate with $\chi^{2}$/d.o.f = 2926/49 and 2630/54 for Obs-1 and Obs-2 respectively. The fractional variability analysis supports these results suggesting the presence of moderate X-ray variability with the amplitude of $F_{\rm var,obs1}$=12.4$\pm0.3\%$ and $F_{\rm var,obs2}$=11.8$\pm0.2\%$. Then we divide the $0.6-10\kev$ light curve into two energy bands: $0.6-2\kev$ and $2-10\kev$ which are the representatives of the soft excess and hard continuum respectively. Figure~\ref{lc} also shows 
the corresponding hardness ratio (HR=$2-10\kev$/$0.6-2\kev$). The fractional variability in the soft and hard bands is similar  with $F_{\rm var,soft}$=13.3$\pm0.4\%$ and $F_{\rm var,hard}$=11.8$\pm0.5\%$ for Obs-1 and $F_{\rm var,soft}$=12.8$\pm0.3\%$ and $F_{\rm var,hard}$=12.7$\pm0.4\%$ for Obs-2. 

To investigate further, we derive the X-ray ($0.6-10\kev$) fractional RMS spectrum which is an important model-independent approach to study the energy-dependent variability in AGN. The high signal-to-noise ratio of \suzaku{} spectra provides an opportunity to understand the variability of the broadband data through RMS spectra (e.g.~\citealt{2003MNRAS.345.1271V}). This mostly model-independent technique can complement the time-averaged spectral analysis and discriminate between models.  
This way we can isolate variable spectral components and study their properties independent of non-variable emission regions. 
By definition, the fractional variability amplitude is the square root of the normalized excess variance $\sigma_{\rm NXS}^2$ \citep{2002ApJ...568..610E}:

\begin{equation}
 F_{\rm var}=\sqrt{\sigma_{\rm NXS}^2}=\sqrt{\frac{\sigma_{\rm XS}^{2}}{\overline{x}^2}}
 \label{eu1}
\end{equation}
\begin{equation}
 \sigma_{\rm XS}^2=S^2-\overline{\sigma_{\rm err}^2}
\end{equation}
where $S^2$ is the sample variance and $\overline{\sigma_{\rm err}^2}$ is the mean square error which is defined by

\begin{equation}
\overline{\sigma_{\rm err}^{2}}=\frac{1}{N}\sum\limits_{i=1}^{N}\sigma_{\rm err,i}^{2}
\end{equation}
The fractional RMS spectrum of 1H~0323+342 is obtained by generating background subtracted light curves in 18 energy bands using the simultaneous and equal length combined XIS0+1+3 data with time bins of $3500\s$ for each observation. The error on $F_{\rm var}$ is estimated in conformity with that estimated by Vaughan et al.2003. The results are shown in Figure~\ref{rms}. For both observations, the source shows an overall decrease in fraction variability with energy while for Obs-2, the fractional rms increases significantly in the range $2-6\kev$.  

To probe the nature of variability above 10\kev{}, we derive the $3-50$\kev{} \nustar{} fractional RMS spectrum of the source. We extract the background subtracted FPMA and FPMB light curves in 9 energy bands with a timing resolution of $\Delta t=3500\s$ and then compute the fractional rms in each energy band from the combined FPMA+FBMB light curves using equation~\ref{eu1}. The resulting fractional rms variability spectrum is shown in Figure~\ref{rms}. To study the energy dependence of variability, we fit a constant model to the $3-50$\kev{} rms spectrum which provides for a statistically acceptable fit with $\chi^{2}$/d.o.f = 5.8/8. The best-fit value of the fractional rms is $13.3\pm1.0$~percent.

\section{discussion}

In this paper, we have performed a broadband spectral analysis of the RLNLS1 galaxy 1H~0323+342 using two simultaneous \suzaku{} and \swift{} 
observations. We have studied the multiwavelength data to shed light on the origin and nature of different components in the spectral energy distribution. The role of disc emission, reflection and thermal Comptonization in a radio-loud AGN is examined in detail. Thermal Comptonization of disc photons in a hot corona is found to be responsible for the primary X-ray continuum which dominates the observed X-ray emission. The presence of prominent soft X-ray excess emission and a broad but weak Fe emission line is confirmed in the source spectrum. We find a possible presence of contribution from the jet in the optical/UV band. Addition of a simple powerlaw produces a satisfactory fit. The $0.3-10 \kev$ fractional RMS variability spectrum is produced to overcome the energy spectral model degeneracy, as it connects the energy spectrum with temporal variability. Further discussion is in order.  

\subsection{Origin of soft excess and Fe $K_{\alpha}$line}

The origin of the soft excess is still ambiguous in nature and consequently RLNLS1 galaxies with prominent soft excess spark immense interest from a scientific point of view.
Both Obs-1 and Obs-2 show the presence of prominent soft excess below $1\kev$. A simple bbody model can describe this excess with $kT\sim0.14\kev$ for both Obs-1 and Obs-2. This is in agreement with \cite{2015AJ....150...23Y}. In other NLS1s and quasars, it varies within a range of $100-200\ev$~\citep{2004MNRAS.349L...7G,2006MNRAS.365.1067C,2007ApJ...671.1284D,2011ApJ...727...31A}. 
The excess at the soft X-ray band along with residuals at around $6\kev$ are commonly seen in radio-quiet (RQ) NLSy1 galaxies. The reflection model {\tt relxill} provides a good fit to the broadband spectrum and indicates a rapid spinning ($a\geq0.9$) black hole.  It suggests blurred reflection from a geometrically thick, moderately ionised accretion disk ($log\xi\geq2$) extending up to $1.2r_{g}$. This is in agreement with the theory that high spin may be necessary for jet
production in AGNs (see e.g., ~\citealt{1977MNRAS.179..433B,2012MNRAS.419L..69N,2013ApJ...762..104S} and others). A MCMC analysis is done to search in the high-dimensional parameter space and check the uncertainties on model parameters. Results show similar probability density and support our finding as global best-fit of the {\tt relxill} model. We also note that this model fails to describe the observed flux at energies above $50\kev$ possibly due to the contribution from jet emission in the hard X-ray spectrum. This scenario enables us to study the connection between the emission from the X-ray corona
and the jet, as has been previously suggested (e.g.,~\citealt{2005ApJ...635.1203M}). 
 The total coronal luminosity found from our spectral modelling is $\sim10^{43.9}\lunit$ for Obs-1 which is in agreement with~\cite{2014ApJ...789..143P} who found the jet kinetic energy and the accretion disk luminosity to be $>10^{45}\lunit$ and $\sim10^{45}\lunit$ respectively. 

\nustar{} data are unable to constrain the parameters of the {\tt relxill} model. We obtain similar results when we use \suzaku{} data above $3\kev$. This proves that the reflection parameters are mainly constrained by the soft excess. The intrinsic disc Comptonization model, {\tt optxagnf} also fit the soft excess but results in excess in the optical/UV and also in the hard X-ray emission above $30\kev$. Simultaneous \swift{} data is used for both Obs-1 and Obs-2. The addition of a broken powerlaw improves the fit significantly. The best-fit parameters indicate a high Eddington ratio($\sim0.45$) and a rapid spinning of the black hole($\sim0.9$) for both observations(See Table~\ref{optxagnf bknpo fit}) which are in agreement with \cite{2014ApJ...789..143P}. Previously \cite{2015AJ....150...23Y} studied the time lag and fractional variability in X-ray and optical/UV band and their results suggested that soft excess might have its contribution from the optically thick thermal Comptonization or the blurred reflection. Our analysis rather shows that the combined reflection and disc Comptonized model provides a satisfactory fit to both observations and supports the possibility that the soft excess may not be entirely due to blurred reflection or thermal Comptonization alone and maybe a combination of both. 

Previous studies of 1H~0323+342 have revealed some contradicting results regarding the Fe emission line. \cite{2013MNRAS.428.2901W} studied the Obs-1 and predicted a highly rotating black hole, but no statistically compelling narrow iron emission lines were detected. \cite{2015AJ....150...23Y} found a systematic excess in the residuals above $6\kev$, and the spin parameter was found to be < 0.13 at the 90\% confidence level. \cite{2014ApJ...789..143P} used averaged X-ray spectrum from \swift{} XRT and found a soft excess along with a residual around the energy of the Fe $K_{\alpha}$ line. Our broadband \suzaku{} data enables us to investigate this feature in detail and both the broad Gaussian line profile and the {\tt RELLINE} model provides a good fit for all three observations. We notice that a narrow Gaussian line profile instead of a broad one results in a significant increase in $\chi^{2}$. Partial covering absorber models (ionised or neutral) which may result in a similar spectrum provides for a poor statistics compared to reflection models. The simultaneous spectral fit of all three observations indicates similar results and the low value of F-test probability supports the presence of a broad Fe emission line. The value of UV spectral slope ($\alpha_{UV}$ = 1.9 and 2.1 for Obs-1 and Obs-2 respectively) also indicates a blue optical/UV spectrum and is in agreement with \cite{2010ApJS..187...64G}. This result further supports that the X-ray spectrum is not affected by a pc absorber. So, we conclude that the emission line is broad in nature but not strong enough to constrain the model parameters. The physical reflection model {\tt relxill} also provided a good fit for all three observations and supports this explanation.

\subsection{UV emission and the jet contribution}

The simultaneous UV to hard X-ray broadband spectral data sets provide us with the unique opportunity to study the UV emission in detail. The Eddington ratio decreases from $\sim 0.74$  to 0.16 between the two \suzaku{} observations. The best-fit electron temperature $kT_{e}$ has a lower limit of $0.37 \kev$ and the optical depth $\tau$ varies within $7.1-13.1$. These results are in agreement with other RLNLSy1 galaxies. The {\tt optxagnf} model combines all the colour-corrected disc emissions from $r_{out}$ to $r_{corona}$ to describe the big blue bump. In our case, the {\tt optxagnf} model fails to describe the UV emission entirely as the intrinsic disk emission for both the \suzaku{}+\swift{} observations (See Sec 5.3). The excess UV emission is consistent with a steep powerlaw with $\Gamma \sim 2.5$.
This, in turn, supports the existence of the contribution from the jet in UV band. Similar results have been found by~\cite{2016MNRAS.460.1705M} and \cite{2015A&A...574A.121K}. They found excess emission in the UV band of the RLNLS1 galaxies RXJ 1633.3+4719 and RX J2314.9+2243. They attributed it to the jet emission. Earlier, the similar analysis of another RLNLS1 galaxy, PKS 0558--504 \citep{2016MNRAS.456..554G}, showed no such excess in the UV band and optical/UV emission is found to be dominated by the disc emission. 
 It should be noted that the assumption of the optxagnf model that all the energy produced by accretion powers the optical/UV and X-ray is not valid for a source with a strong jet. Hence the Eddington ratio $L/L_{Edd}$ inferred from the optxagnf model should refer to the disk/corona emission only, and should be considered a lower limit to the actual $L/L_{Edd}$ that includes the jet contribution also.

\begin{figure}
\includegraphics[width=7.5cm,height=8.5cm,angle=-90]{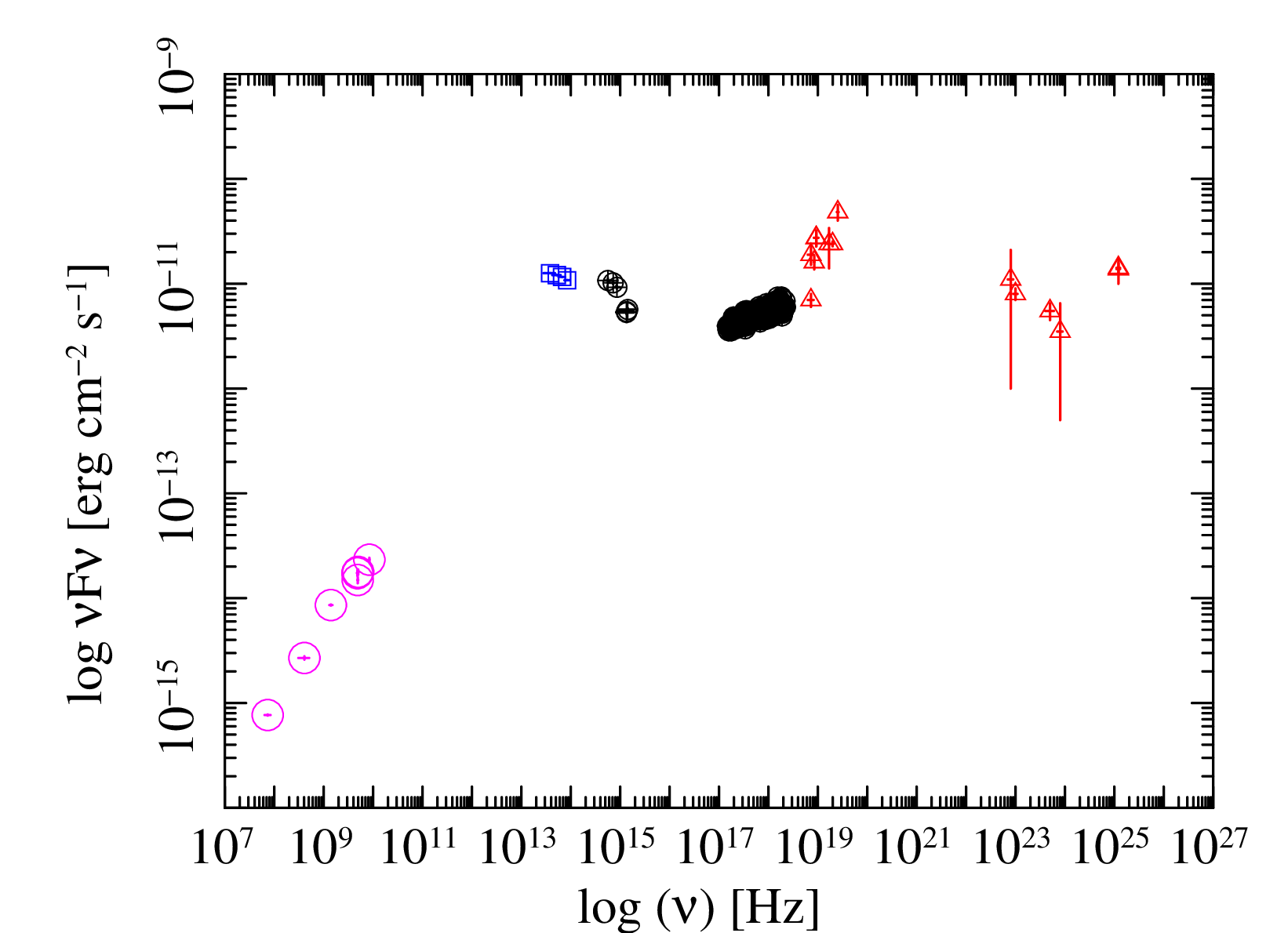}
\caption{The the SED of 1H 0323+342 using Obs-2 Optical/UV to hard X-ray data along with detections in the hard X-ray to $\gamma$-ray by Swift/BAT~\citep{2013ApJS..207...19B} and INTEGRAL/ISGRI~\citep{2007ApJS..170..175B} and Fermi/LAT 2 year Point Source Catalog~\citep{2012ApJS..199...31N} respectively. Archival radio data are from NED. The radio, infrared, simultaneous optical-UV (\swift{} UVOT) to hard X-ray(\suzaku{} XIS and PIN) and Hard X-ray/GeV data are plotted in magenta(circles), blue(boxes), black(crossed circles) and red(tri angles) respectively. See the electronic edition of the Journal for a color version of this figure.}
\label{sed}
\end{figure}

The overall SED from the radio to the Gamma-rays can provide useful information about the contribution from the jet and the 
accretion disk. Simultaneous observation at different energy band is ideally required to construct the SED due to a relatively large variability of this source. We have used archival but not simultaneous data to plot the SED which is shown in Figure~\ref{sed} where Optical/UV to hard X-ray data of Obs-2 is included. The radio data are taken from the NASA/IPAC Extragalactic Database(NED). Hard X-ray and $\gamma$-ray data are included from the Swift/BAT 70 Month Catalog~\citep{2013ApJS..207...19B}, INTEGRAL/ISGRI~\citep{2007ApJS..170..175B} and the Fermi/LAT 2 year Point Source Catalog~\citep{2012ApJS..199...31N} respectively. It is hard to draw any correct conclusion because the data sets are not simultaneous but the jet emission seems to make a noticeable contribution to the hard X-ray band (>10 keV) and surely dominates the observed flux at energies around 100 keV. This is supported by our Swift-BAT analysis where we have studied the BAT light curve of this AGN but a $\chi^{2}$ test has not been possible as the error bar is consistent with zero. We have compared the BAT spectrum with the \nustar{} and the \suzaku{} HXD data and they show similar trends with respect to the BAT spectrum. A simple powerlaw model fits the hard X-ray spectra (above $10 \kev$) well with $\chi^{2}/dof = 269/217$ and reveals an excess around $100 \kev$ as found in Figure~\ref{sed}. We note that {\tt optxagnf}, results in a significant excess in the UV band but successfully describes observed optical, UV and X-ray continuum spectra for radio-quiet AGNs where emission from the disc dominates. We have used the peak flux ($\sim1.1\times10^{-11} \funit$/\AA) at the \swift{} UVOT band which provides a monochromatic luminosity at $8\times10^{14}$ Hz or at $3747$\AA. We have calculated this monochromatic luminosity to be $\sim9\times10^{43} \lunit$/\AA{} for Obs-2 which is almost half of the disk luminosity and $\sim 10\%$ of the $L_{EDD} (1.38\times10^{45}\lunit)$ for a $10^{7} \msol$ BH. The disc luminosities are calculated from the best-fit of our \swift{} UVOT data and are found to be $L_{disc}\sim 2.10\times10^{44}\ergs$ for Obs-1 and $\sim 1.82\times10^{44}\ergs$ for Obs-2 respectively. We have used {\tt diskbb} to fit the six UVOT data points and calculate the disc luminosity with the assumption that this emission is entirely from the disk. If we neglect the host galaxy contamination, the excess UV emission can be due to the high energy tail of synchrotron emission from a jet. So, we conclude that the contribution from the jet is significant in the UV energy band during the time epoch studied by us. 

The bolometric luminosity is found to be $L_{bol} = 6.26\times10^{44}\lunit$ and $4.89\times10^{44}\lunit$ for Obs-1 and Obs-2 respectively. Here we calculate the unabsorbed flux in the energy band $0.001 - 100\kev$ using the convolution model {\tt cflux}. The Eddington ratio for this radio-loud AGN is = $\frac{\dot{M}}{\dot{M_{\rm E}}}$ = $\frac{L_{bol}}{L_{E}}$ = 0.45 and 0.35 for Obs-1 and Obs-2 respectively. This is in agreement with our spectral best fit and with previous and recent studies~\citep{2012AstL...38..475K,2017MNRAS.464.2565L}. 
  The $2-10\kev$ unabsorbed power-law luminosities are found to be $\sim 7.94\times10^{43}\ergs$for Obs-1 and $\sim 5.01\times10^{43}\ergs$ for Obs-2 respectively and are calculated using the optical/UV and X-ray data of the two observations. The ratio of $L_{disc}$ and power-law luminosities provide an estimation of the accreted energy dissipated into the accretion disk and the hot corona. Hence, approximately 38 (Obs-1) to 27 (Obs-2) percent of the accreted energy is used to fuel the hot corona in 1H~0323+342. The amount is likely to increase if we take in to account the jet contribution in the UV band.

\subsection{Variability significance}

Previously, both long-term and short-term variability has been found in 1H~0323+342~\citep{2015AJ....150...23Y}. Our fractional variability analysis supports these results, suggesting a moderate X-ray variability with the amplitude of $F_{\rm var,obs1}$=12.4$\pm0.3\%$ and $F_{\rm var,obs2}$=11.8$\pm0.2\%$ for Obs-1 and Obs-2, respectively. The variabilities in both soft and hard bands are almost roughly the same with $F_{\rm var,soft}$=13.3$\pm0.4\%$ and $F_{\rm var,hard}$=11.8$\pm0.5\%$ for Obs-1 and $F_{\rm var,soft}$=12.8$\pm0.3\%$ and $F_{\rm var,hard}$=12.7$\pm0.4\%$ for Obs-2. 
The fractional RMS spectrum of 1H~0323+342 for both \suzaku{} observations show enhanced variability below $1\kev$ and remain almost constant $F_{var}$ above $3\kev$. Due to large error bars at higher energies, it is difficult to infer any trend and we can only argue that both soft excess and powerlaw components are variable in nature.

\subsection{Broadband Continuum}

Our simultaneous UV to hard X-ray broadband spectral fitting reveals the presence of prominent soft excess below $1\kev$, variable primary emission in the hard X-ray band above 2 keV and a contribution from the jet in the UV. We find a weak but broad Fe emission line and a weak reflection hump in all three observations. Our spectral modelling predicts the disc to extend up to the innermost regions and suggests an ionized disc, although we note that the blurred reflection best-fit parameters are constrained by soft excess only and can be a combined effect of blurred reflection and thermal Comptonization. From the modelling of the broadband spectrum, we estimate the relative contribution of different components to the observed flux in the UV and the X-ray bands. The $0.6-10\kev$ flux indicates a decrease(from $1.9\times10^{-11}$ to $1.6\times10^{-11}$) between the two observations which correspond to $L_{X}\sim 1.5\times10^{44}\lunit$ and $1.25\times10^{44}\lunit$. Our finding is in agreement with \cite{2015AJ....150...23Y} for Obs-1. The flux in the UV band has increased from $2.36^{+0.01}_{-0.01}\times10^{-11}$ to $2.46^{+0.01}_{-0.01}\times10^{-11}$ between the two observations. This result supports our spectral analysis where the accretion rate has gone lower and powerlaw norm, which represents the UV flux, has increased.

\section{Conclusions}

The main results of our detailed analysis of optical/UV and X-ray emission from 1H~0323+342 based on \swift{}, \suzaku{} and \nustar{} are as follows.

\begin{enumerate}
	\item The optical-to-hard X-ray spectrum of 1H~0323+342 is complex, and consists of multiple spectral components -- a broadband X-ray power-law ($\Gamma \sim 1.8$), an iron line and hard X-ray excess emission, described by a blurred reflection component ($R\sim 0.5$), a soft X-ray excess component, strong optical/UV emission.

   \item The strong UV emission cannot be accounted  by an accretion disk alone, the excess UV emission is in the form of a steep power law ($\Gamma \sim 3-3.5$), which most likely arises from the jet in the radio-loud NLS1. 

	\item The iron line around $6.4\kev$ is weak but broad in nature, and along with the hard X-ray excess emission, consistent with the blurred reflection component with $R\sim 0.5$.

	\item The soft X-ray excess is well described by both blurred reflection and thermal Comptonization and it may well, have a contribution from both the processes.

	\item The rapid spin of the black hole and the inner extent of the accretion disk indicate that truncated disk and retrograde spin may not be necessary for launching the jets.

         \item Similar to the radio-quiet NLS1, both the soft X-ray excess and the powerlaw components are variable in nature. In the $3-50\kev$ band, the variability does not appear to depend on energy,  and likely caused by the variations in the X-ray powerlaw normalisation.

\end{enumerate}

\section{Acknowledgements}
We thank the anonymous referee for many useful comments and
suggestions which helped to improve the quality of the manuscript.
RG acknowledges the financial support from  Visva-Bharati University and IUCAA
 visitor programme. BR likes to thank IUCAA for their facilities and hospitality provided
to him during his visits under their Visiting Associateship
programme. This research has made use of  \suzaku{}, \swift{} and \nustar{} data, software and/or web tools obtained from NASA's High Energy Astrophysics Science Archive Research Center (HEASARC), a service of Goddard Space Flight Center and the Smithsonian Astrophysical Observatory. This research has made use of the \nustar{} Data Analysis Software (NuSTARDAS) jointly developed by the ASI Science Data Center (ASDC, Italy) and the California Institute of Technology (USA). We also thank the UVOT calibration team for providing useful software.

\bibliographystyle{mnras} \bibliography{mybib}

\end{document}